\documentclass[secnumarabic,amssymb, nobibnotes, aps, prd]{revtex4-2}
\usepackage{graphicx}
\usepackage{amsmath}
\usepackage{subcaption}
\usepackage{xcolor}
\usepackage{multirow}

\begin{document}
	
	\title{Topology of restricted phase space thermodynamics in  Kerr-Sen-Ads black holes }%

	\author{Bidyut Hazarika$^1$}
	
	\email{$rs_bidyuthazarika@dibru.ac.in$}
	
	\author{Prabwal Phukon$^{1,2}$}
	\email{prabwal@dibru.ac.in}
	
	\affiliation{$1.$Department of Physics, Dibrugarh University, Dibrugarh, Assam,786004.\\$2.$Theoretical Physics Division, Centre for Atmospheric Studies, Dibrugarh University, Dibrugarh, Assam,786004.}
	\maketitle
	\section*{Abstract}
	In this study, we investigate the thermodynamic topology of the Kerr-Sen-Ads black hole in restricted phase space. In the restricted phase space, a new parameter, central charge $C$, and its conjugate parameter $\mu$ are introduced, omitting the well-known $PdV$ term in the first law of black hole thermodynamics. We study the local and global topology of the black hole by considering the black hole solution as topological defects in the free energy landscape. We compute the winding number and the total topological number at the thermodynamic defects. For our analysis, we have considered five ensembles of Kerr-Sen-Ads black holes in restricted phase space: fixed $(Q, J, C)$, fixed $(\phi, J, C)$, fixed $(Q,\Omega, C)$, fixed $(Q, J, \mu)$, and fixed $(\phi,\Omega, C)$, where $Q$ is the electric charge, $J$ is the angular momentum, $C$ is the central charge, $\phi$ is the electric potential conjugate to charge, $\Omega$ is the angular frequency conjugate to $J$, and finally, $\mu$ is the chemical potential. In the fixed $(Q, J, C)$, fixed $(\phi, J, C)$, and fixed $(Q, J, \mu)$ ensembles, we find a topological charge of $+1$. In the fixed $(Q,\Omega, C)$ and fixed $(\phi, \Omega, C)$ ensembles, depending on the values of the thermodynamic parameters, we find topological charges of $-1$, $0$, and $+1$. Interestingly, in ensembles where we find the topological charge to be $0$, we observe both Hawking-Page and Davies type phase transitions. We show that both types of these phase transitions can be studied using a common vector field, and the topological charges associated with Davies type and Hawking-Page phase transitions are $-1$ and $+1$, respectively. 
	\section{Introduction:}

The study of black hole thermodynamics has made substantial strides since its inception in the 1970s, largely attributed to the groundbreaking work of Bekenstein and Hawking, who established the link between thermodynamic quantities and black hole physics \cite{Bekenstein:1973ur,Hawking:1974rv,Hawking:1975vcx}. Bekenstein introduced the revolutionary concept of black hole entropy, while Hawking's discovery of black hole radiation demonstrated that black holes adhere to thermodynamic laws, fundamentally establishing that black holes are governed by thermodynamic principles \cite{Bardeen:1973gs}. Since these foundational discoveries, numerous advancements have deepened the understanding of this intricate relationship \cite{Wald:1979zz,bekenstein1980black,Wald:1999vt,Carlip:2014pma,Wall:2018ydq,Candelas:1977zz,Mahapatra:2011si}. A particularly important insight in the mechanics of black holes is the occurrence of phase transitions\cite{Davies:1989ey,Hawking:1982dh,curir_rotating_1981,Curir1981,Pavon:1988in,Pavon:1991kh,Kastor:2009wy,Dolan:2010ha} which were first identified by Davies \cite{Davies:1989ey}. This critical finding showed that black hole phase transitions are marked by a discontinuity in the heat capacity at certain points.
Another significant type of phase transition is the Hawking–Page transition \cite{Hawking:1982dh}, which occurs when the sign of the black hole’s free energy changes. Transitions from non-extremal to extremal black hole states have also been extensively explored in various works \cite{curir_rotating_1981,Curir1981,Pavon:1988in,Pavon:1991kh}. Additionally, many studies have investigated phase transitions that resemble those found in van der Waals systems, further enriching the understanding of black hole thermodynamics \cite{Kastor:2009wy,Dolan:2010ha,14,15}.\\

The study of black hole phase transition behaviour has been given a new dimension through the introduction of extended phase-space thermodynamics \cite{14,15,new}. This framework introduces a novel pair of variables (P, V) into the domain of thermodynamic parameters, where $P$ and $V$ represents pressure and volume respectively.  The pressure $P$ can be expressed in terms of the negative cosmological constant$\Lambda$ as: $$P=\frac{-\Lambda}{8 \pi G}$$
The first law in the EPST can be written as
$$M = T dS+\Omega dJ+ \Phi dQ +V dP$$
where $M$, $T$, and $S$ respectively represent the mass, temperature, and entropy of the black hole. $J$ and $\Omega$ are the angular momentum, and its conjugate parameter angular velocity while $Q$ and $\phi$ are the electric charge and its conjugate parameter Coulomb potential. The Smarr relation in EPST holds the form

$$M = \frac{d-2}{d-3}\left(TS+ \Omega J\right) +\Phi Q-\frac{2}{d-3}PV$$

One key issue within the EPST framework arises when allowing $P$ the pressure, to vary, as this affects the cosmological constant which is a fundamental parameter that defines the gravitational theory. Therefore, altering $P$ implies changing the underlying gravity model, meaning that the ensemble no longer represents black holes in the same gravitational framework but instead considers different gravity models that yield similar black hole solutions. This is the famous so-called "ensemble of theories" problem.EPST has several issues such as the presence of the \(V dP\) term, which suggests that the black hole mass \(M\) should be viewed as enthalpy rather than internal energy which makes the physical meaning of the thermodynamic volume \(V\) ambiguous. One of the
most significant issues arises when comparing the first law to the Smarr relation. By differentiating the mass $M$ with respect to the entropy $S$, angular momentum $J$, electric charge $Q$, and pressure$P$, we obtain expressions for the temperature $T$, angular velocity $\Omega$, electric potential \(\Phi\), and thermodynamic volume \(V\). Substituting these relations into the Smarr formula shows that the mass \(M\) behaves as a homogeneous function of these variables. with different orders which depend upon $d$. However, this contradicts standard thermodynamic principles, where the potentials should be homogeneous functions of the extensive variables of the first order.\\

Inspired by Maldacena's renowned work on the AdS/CFT correspondence \cite{Maldacena:1997re}, Visser recently addressed this issue. \cite{H4} by introducing the central charge \(C = \frac{l^{d-2}}{G}\) from the dual CFT theory as a new extensive variable and by considering mass \(M\) as a function of \(A\) (the area of the event horizon), \(J\), \(Q\), \(\Lambda\), and \(G\). This leads to a modified first law:

$$
dM = \frac{\kappa}{2\pi} d\left(\frac{A_{d-2}}{4G}\right) + \Omega dJ + \frac{\Phi}{l} d(Ql) - \frac{M}{d-2} \frac{dl^{d-2}}{l^{d-2}} + \ldots,
$$

where $l$ is the AdS radius, $\kappa$ is the surface gravity, and $G$ is the gravitational constant. In dual CFT, the first law to be rewritten as : 
$$dE = TdS + \Omega dJ + \tilde{\Phi} d\tilde{Q} - PdV + \mu dC$$
where \( \mu \) is the chemical potential conjugate variable to the central charge \( C \). \( \mu \)  signifies the number of microscopic degrees of freedom in the dual CFT.
Visser's formulation avoids the "ensemble of theories" issue by restoring the internal energy role of \(E\), but varying \(V\) still implies changing the cosmological constant \(\Lambda\), thus leaving some unresolved challenges related to the "ensemble of theories" issue .\\

A more specialized version of this framework, known as the Restricted Phase Space Thermodynamics (RPST), was later introduced for AdS black holes \cite{rp0}. This formalism restricts the phase space, focusing on a subset of thermodynamic variables while preserving essential features of black hole thermodynamics. RPST has been applied to various black hole systems \cite{rp1, rp2, rp3, rp4, rp5, rp7, rp8, rp9, kerrsen, rp10, rp11,rp12} and has the potential to enhance our understanding of the AdS/CFT correspondence through black hole thermodynamics. The introduction of Restricted Phase Space Thermodynamics (RPST) provides an alternative by restricting the variation of $P$ and $V$ term which eventually simplifies the first law as ;
\[
dM = TdS + \Omega dJ + \tilde{\Phi} d\tilde{Q} + \mu dC,
\]
Here AdS length \(l\) is kept fixed which also excludes volume work, thereby simplifying the analysis of thermodynamic processes for black holes.
The associated Euler relation gets modified as :
\[
M = TS + \tilde\Phi \tilde Q + \Omega J + \mu C.
\]

These equations follow the typical structure of the first law and Euler relation found in classical thermodynamics. Since the RPST framework allows for the gravitational constant \(G\) to vary, it is important to clarify the different roles of varying \(G\) versus varying the cosmological constant \(\Lambda\). Essentially, \(\Lambda\) is part of the Lagrangian density, so changing \(\Lambda\) results in a different set of gravitational field equations. In contrast, \(G\) appears as an overall factor in front of the total action, ensuring that changing \(G\) does not alter the field equations.  Another distinction is that varying \(\Lambda\) alters the spacetime geometry while varying \(G\) leaves it unchanged.

In addition to the extended phase space and RPST formalisms, several other approaches, such as holographic thermodynamics \cite{H1,H2,H3,H5} and geometric interpretations of phase transitions \cite{geo1,geo2,geo3,geo4,Gogoi:2023qni,Kumar:2012ve}, have been developed to explore critical phenomena in black hole thermodynamics.

A novel avenue in the context of understanding critical phenomena in black hole thermodynamics is the incorporation of topology in black hole thermodynamics initiated by the authors of \cite{28,29}. Duan's pioneering contributions to the study of relativistic particle systems\cite{g1,g2} serve as a key inspiration for this work. This framework emphasizes the zero points of a vector field and signifies the critical points of the system. These zero points serve as indicators of phase transitions, characterized by winding numbers. This topological approach has gained widespread acceptance in various black hole models, as seen in the literature \cite{t1,t2,t3,t4,t5,t6,t7,t8,t9,t10,64, newfield}. The most effective
 topological framework to understand black hole thermodynamics was proposed in \cite{29}, which is particularly useful for classifying black holes into some topological class. By employing the off-shell free energy method, black holes are conceptualized as topological defects within their thermodynamic landscape. This perspective illuminates both the local and global topological attributes of black holes, with their stability and topological charge described through winding numbers. The sign of a black hole's winding number indicates its stability. This approach has proven effective in studying a diverse array of black hole systems across various gravitational theories \cite{t11,t12,t13,t14,t15,t16,t17,t18,t19,t20,t21,t22,t23,t24,t25,t26,t27,t28,t29,t30,wu1,wu2,wu3,wu4,wu5,wu6,wu7,wu8}.
The following paragraph provides a brief overview of the off-shell method and outlines the procedure for computing the topological charge.\\
	
	The expression for the off-shell free energy of a black hole with arbitrary mass is given by\cite{29}
	\begin{eqnarray}
		\mathcal{F}=E-\frac{S}{\tau} \label{8}  
	\end{eqnarray}
	Here, $E$ and $S$ are the energy and entropy of the black hole respectively. $\tau$ is the time scale parameter which equals to :
	\begin{equation}
		\tau=\frac{1}{T}
	\end{equation}
	Here,  $T$ is the equilibrium temperature at the surface of the cavity that encloses the black hole. The time parameter $\tau$ is set freely to vary instead of the mass. Utilizing the generalized free energy, a vector field is constructed as follows \cite{29} :
	\begin{eqnarray}
		\phi=\left(\phi^S,\phi^\Theta \right)=\left(\frac{\partial\mathcal{F}}{\partial S},-\cot\Theta ~\csc\Theta  \right)
	\end{eqnarray}
	 Where $\theta=\pi/2$ and $\tau=1/T$ represent a zero point of the vector field $\phi$. In the $\theta$ component, a particular combination of trigonometric functions is chosen because at $\theta=\frac{\pi}{2}$, the vector field diverges, allowing one to easily identify the zero point of the $S$ component by simply examining the vectors along the $\theta=\frac{\pi}{2}$ line. While constructing the vector field, Duan's $\phi$ mapping current theory has been taken into consideration. Using Duan's theory, the topological current can be written as \cite{28,g1}
	\begin{equation}
	j^\mu=\frac{1}{2 \pi} \epsilon^{\mu \nu \rho} \epsilon_{a b}\partial_\nu n^a\partial_\rho n^b ;\hspace{0.8cm} \mu,\nu,\rho=0,1,2
	\end{equation}
	Here $\partial_\nu=\frac{\partial}{\partial x^\nu}$.  $\epsilon$ is the Levi Civita symbol and $n^{a,b}$ are the normalized vectors.It is important to note that the topological current $j^\mu$ is conserved\cite{g1} i.e
	\begin{equation}
	\partial_\mu j^\mu=0
	\end{equation}
The normalized vector is calculated as :
	\begin{equation}
		n^a=\frac{\phi^a}{||\phi||}; \quad a=1,2  \quad \text{with} \quad \phi^1=\phi^{S}, \quad \phi^2=\phi^{\theta}.
	\end{equation}
	The two conditions that has to be satisfied by $n^a$ are \cite{g1} :
\begin{equation}
	n^an_a=1 \quad \text{and} \quad n^a\partial_\nu n_a=0.
\end{equation}
To calculate the topological charge we use the residue method \cite{64}. This method involves first determining the winding number $(w_{i})$ for each pole using the residue theorem and by adding the winding numbers for each pole, the topological charge can be computed. 
Initially, we find a solution for $\tau$ by solving the equation:
\begin{equation}
	\frac{\partial\mathcal{F}}{\partial S}=0
	\label{sol}
\end{equation}
Let us consider the solution of eq.\ref{sol}, is found to be $\tau(S)$, we construct a function $\mathcal{G}(z)$ as follows ;
$$\tau(S \leftrightarrow z)=\mathcal{G}(z)$$
We aim to create a two-dimensional space where we can introduce a real contour. This contour will encircle the thermodynamic imperfections described by equation \ref{sol}.  Subsequently, a rational complex function $\mathcal{R}(z)$ is constructed using $\mathcal{G}(z)$  as follows\cite{64}
\begin{equation}
	\mathcal{R}(z,\tau)=\frac{1}{\tau - \mathcal{G}(z)} \label{residue2}
\end{equation}

The complex function $R(z)$ is analytic except for a limited number of specific points $z_1,z_2,z_3.....z_m$. To enclose these points, a real closed contour $C$ is introduced where the complex function is analytic within and on the contour. This is followed by employing the residue theorem to compute the integral of the function $R(z)$ over the path $C_i$ surrounding $z_i$ or
\begin{equation}
	\oint \frac{R(z)}{2 \pi i} dz=\sum_{k=1}^{m} Res[R(z_k)]
\end{equation}  

Finally, local topology can be studied by computing the winding number at thermodynamic defects. We do that by just taking normalized  residues at the poles of $\mathcal{R}(z)$ 
\begin{eqnarray}
	w_{i}=\frac{Res\mathcal{R}(z_{i})}{|Res\mathcal{R}(z_{i})|}=Sign [Res\mathcal{R}(z_{i})]\label{residue1}
\end{eqnarray}
Again, the Cauchy-Goursat theorem says that the integral around contour $C$, winding around the singular points, is equal to the integrals of $R(z)$ along $C_i$ which $z_i$ is enclosed. Applying this concept we can either consider a bigger contour that encompasses all thermodynamic defects or we can simply add the individual winding numbers $w_i$, computed around small individual contours. Hence the total winding number or topological charge, $W$, is then computed as $$W=\sum_{i} w_{i}$$\\
In this study, we investigate the thermodynamic topology of Kerr-Sen-Ads black holes in restricted phase space. We consider the black hole solution as a topological defect in the thermodynamic phase space.  To study the local and global topology of the black hole,  we calculate the winding number and the total topological charge at the thermodynamic defects. 	Incorporating the Restricted Phase Space Thermodynamics (RPST) formalism introduces four additional ensembles, resulting in a total of eight: fixed \((Q, J, C)\), fixed \((\phi, J, C)\), fixed \((Q, \Omega, C)\), fixed \((Q, J, \mu)\), fixed \((\phi, \Omega, C)\), fixed \((\phi, J, \mu)\), fixed \((Q, \Omega, \mu)\), and fixed \((\phi, \Omega, \mu)\). Here, \(Q\) denotes the electric charge, \(J\) the angular momentum, and \(C\) the central charge, while \(\phi\), \(\Omega\), and \(\mu\) represent the electric potential (conjugate to \(Q\)), the angular velocity (conjugate to \(J\)), and the chemical potential, respectively. The thermodynamic properties of black holes across all these ensembles remain largely unexplored in the literature. In this work, we focus on investigating the thermodynamic topology of the first five ensembles in detail. The choice of ensemble often plays a critical role in shaping the thermodynamic behavior and phase transitions of black holes \cite{ec1, ec2, ec3, ec4, ec5, ec6}. For each of these ensembles, we analyze their thermodynamic topology by computing the corresponding topological charge. For example, in the fixed $(Q,J,C)$ and $(\phi,J,C)$ ensemble, we observed Van der wall like phase  transition, whereas in the fixed $(Q,\Omega,C)$ ensemble, Hawking-Page and Davies type phase transition is observed. Since the thermodynamics of the black hole change with a shift in ensemble, it merits detailed investigation to determine whether the topology of the thermodynamic spaces also changes. We naturally anticipate that the thermodynamic topology of the two ensembles will differ because they have distinct domain parameters that define their respective thermodynamic phase spaces. In summary, as the ensemble changes, the thermodynamic parameters associated with the phase space of the black hole also change, thereby altering its thermodynamic behaviors. Consequently, the associated thermodynamic topology undergoes changes due to shifts in the thermodynamic phase space. This ensemble dependence is precisely what our work investigates.\\
		
		This paper is organized into the following sections: in Section \textbf{II}, we briefly review the Kerr-Sen-Ads black hole in restricted phase space.  We  analyse the thermodynamic topology of fixed ($Q, J, C$) ensemble in subsection \textbf{II.I.}, fixed ($\phi, J, C$) in subsection \textbf{II.II.}, fixed ($Q, \Omega, C$) in subsection \textbf{II.III.} , fixed ($Q, J, \mu$) in subsection \textbf{II.IV.} and finally in subsection \textbf{II.V}, we analyze the same for fixed $(\phi, \Omega, C)$ ensemble. The conclusions are presented in Section \textbf{III}.
	\section{Kerr-Sen Ads black hole in restricted phase space}
The Kerr-Sen Ads black hole is the charged rotating black hole solution to the low-energy limit of heterotic string theory discovered by Sen \cite{ks1}. The Kerr-Sen black holes in four dimensions have been examined from various perspectives\cite{kerrsen,ks2,ks3,ks4,ks5,ks6,ks7,ks8,ks9,ks10}. The thermodynamics of phase space in the extended phase space was studied in \cite{ks3} and the thermodynamics of Kerr-Sen Ads black hole was studied in \cite{kerrsen}. In this section a brief review of the Kerr-Sen-Ads black hole in restricted phase space is provided\cite{kerrsen}.\\
 The metric for the Kerr-Sen-Ads black hole in the Boyer-Lindquist coordinate is given by \cite{kerrsen} :
		\begin{equation}
		ds^2=-\frac{\Delta_r}{\rho^2}\left [ dt-\frac{a sin^2\theta d\varphi}{\Xi} \right ]^2+\frac{\rho^2}{\Delta_r} dr^2 +\frac{\rho^2}{\Delta_\theta} d\theta^2+\frac{\Delta_\theta sin^2 \theta} {\rho^2}\left[a dt-\frac{r^2+a^2}{\Xi} d\varphi\right]^2
		\label{rcmetric}
	\end{equation}
	where 
	$$\Delta_{r}=(r^2+2br+a^2)\left(1+\frac{r^2+2br}{l^2}\right)-2Gmr$$
	$$\Xi=1-\frac{a^2}{l^2} \hspace{0.5cm};\hspace{0.5cm} \rho^2=r^2+2br+a^2cos^2\theta,$$
	$$\Delta_\theta=1-\frac{a^2}{l^2} cos^2\theta$$
The parameter $a$ represents the specific angular momentum of the black hole, defined as the ratio of the black hole's angular momentum $J$ to its mass $M$. For a non-rotating black hole, $a=0$.	 $b$ is the diatonic charge which can be expressed in terms of black hole mass $m$ and electric charge $q$ as $b=q^2/2m.$
	The expressions for ADM  mass $M$ and the angular momentum  $J$ and charge $Q$ are :
	\begin{equation}
		M=\frac{m}{\Xi^2} \hspace{0.5cm};\hspace{0.5cm} 	J=\frac{a m}{\Xi^2} \hspace{0.5cm} \text{and} \hspace{0.5cm} Q=\frac{q}{\Xi}
		\label{eq8}
	\end{equation}
	using the condition $\Delta_r(r=r_+)=0$ we can write $M$ and $J$ as
		\begin{equation}
		M=\frac{1}{2 \Xi^2 G r_+}(r^2+2br+a^2)\left(1+\frac{r^2+2br}{l^2}\right)
	\end{equation}
and 
\begin{equation}
J=\frac{a}{2 \Xi^2 G r_+}(r^2+2br+a^2)\left(1+\frac{r^2+2br}{l^2}\right)
\end{equation}
where $r_+$ is the event horizon radius,.$l$ is the Ads radius, for $l->\infty$, the Kerr-Sen AdS black hole reduces to the usual Kerr-Sen black hole.G is the  Newton's gravitational constant.  To introduce RPST formalism two new quantities are defined which is the central charge $C$ and its conjugate parameter $\mu$\cite{H4,rp0,rp7}.
\begin{equation}
	C=\frac{l^2}{G} ; \hspace{1cm} \mu=\frac{M-T S+\Omega J+\hat{\phi} \hat{Q}}{C}
\end{equation}
Here $\hat{Q}$ and $\hat{\phi}$ is the rescaled charge and electric potential expressed as\cite{rp7}
\begin{equation}
		\hat{Q}=\frac{Ql}{\sqrt{G}};\hspace{1cm} \hat{\phi}=\frac{\phi \sqrt{G}}{l}
\end{equation}
Using these defined quantities we write the two most important equations in RPST formalism. The first one is the modified first law of black hole thermodynamics which is incorporated with a new term $\mu dC$ and made free from the usual $\mathcal{P}d\mathcal{V}$ term : 
\begin{equation}
	dM=TdS+\Omega dJ+\phi dQ+\mu dC
\end{equation}
and followed by the Euler-like relation :
\begin{equation}
	M=TS+\Omega J+\hat{\phi} \hat{Q}+\mu C
\end{equation}
Now, rewritting $a,G$ as follows:
\begin{equation}
		a=\frac{J}{M}; \hspace{1cm} G=\frac{l^2}{C}
		\label{eq22}
\end{equation} 
we can expressed the horizon radius $r_+$ as \cite{kerrsen}:
\begin{equation}
	r_+=\frac{1}{2 M}\left(\sqrt{\frac{-4 J^2(\pi C+S)+\pi C Q^4+4 l^2 M^2 S}{\pi C}}-Q^2\right)
	\label{eq23}
\end{equation}
Putting equation.\ref{eq22} and equation.\ref{eq23} into equation.\ref{eq8}, the ADM mass is obtained as \cite{kerrsen} :
	\begin{equation}
	M=\frac{\sqrt{\pi  C+S} \sqrt{\pi  C \left(4 \pi ^2 J^2+S^2\right)+2 \pi ^2 Q^2 S+S^3}}{2 \pi ^{3/2} l \sqrt{C S}}
	\label{mass}
\end{equation}
rest of the thermodynamic quantities can be computed using the first law of thermodynamics. For details of the RPST formalism of Kerr-Sen-Ads black hole see \cite{kerrsen}.
Now we study the thermodynamic topology in five different ensembles in the following subsections.
	\subsection{Fixed (Q,J,C) Ensemble}
\begin{figure}[h!t]
		\centering
		\includegraphics[width=10cm,height=8cm]{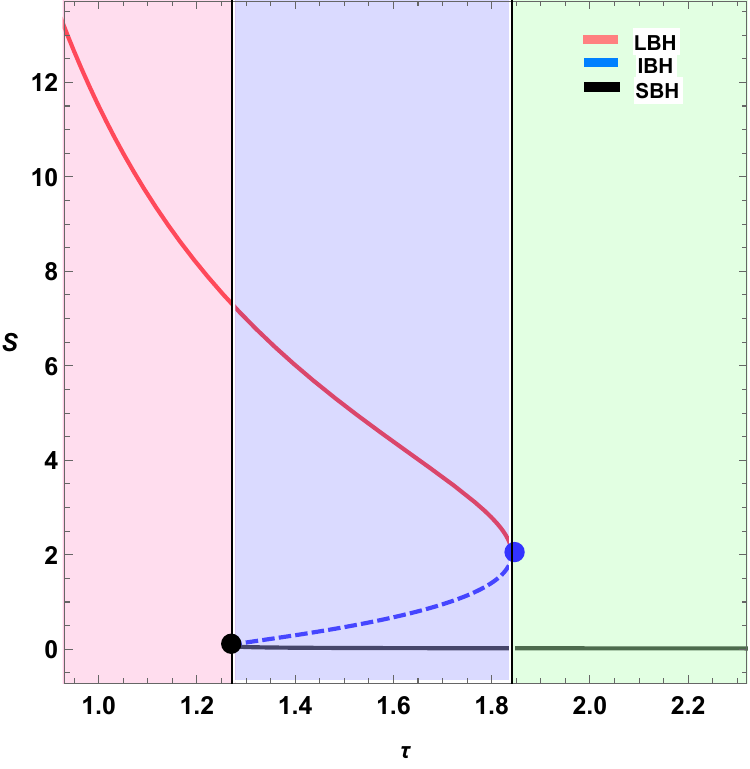}
		\caption{$\tau$ vs $S$ plot for Kerr Sen Ads  black hole at  $C=0.2,Q=1.5,J=0.02$ and $l=1$. The $LBH$ and $SBH$ regions represent the large and small black hole branches respectively.}
		\label{1}
	\end{figure}

	For Kerr-Sen Ads black hole, mass $M$ in this ensemble is  :
	\begin{equation}
		M=\frac{\sqrt{\pi  C+S} \sqrt{\pi  C \left(4 \pi ^2 J^2+S^2\right)+2 \pi ^2 Q^2 S+S^3}}{2 \pi ^{3/2} l \sqrt{C S}}
		\label{mass}
	\end{equation}
	Using eq. \ref{mass},  free energy is computed as :
	\begin{equation}
		\mathcal{F}=M-S/\tau=-\frac{2 \pi ^{3/2} l S \sqrt{C S}-\tau  \sqrt{\pi  C+S} \sqrt{4 \pi ^3 C J^2+\pi  C S^2+2 \pi ^2 Q^2 S+S^3}}{2 \pi ^{3/2} l \tau  \sqrt{C S}}
	\end{equation}
	The  components of the vector field $\phi$ is obtained as :
	\begin{equation}
	\phi^r=-\frac{\alpha}{\beta}
	\end{equation}
	where,
	\begin{multline*}
		\alpha =4 \pi ^4 C^3 J^2 \tau -4 \pi  C^2 S^3 \tau -\pi ^2 C^3 S^2 \tau +4 \pi ^{3/2} l \sqrt{\pi  C+S} (C S)^{3/2} \sqrt{4 \pi ^3 C J^2+\pi  C S^2+2 \pi ^2 Q^2 S+S^3}\\-2 \pi ^2 C Q^2 S^2 \tau -3 C S^4 \tau 
	\end{multline*}
	and
	\begin{equation*}
	\beta={4 \pi ^{3/2} l \tau  (C S)^{3/2} \sqrt{\pi  C+S} \sqrt{4 \pi ^3 C J^2+\pi  C S^2+2 \pi ^2 Q^2 S+S^3}}
	\end{equation*}
	The $\Theta$ component is :
	\begin{equation}
		\phi^\Theta=-\cot\Theta ~\csc\Theta 
	\end{equation}
	The zero points of $\phi^r$ is also obtained as:
	\begin{equation}
		\tau=\frac{4 \pi ^{3/2} l (C S)^{3/2} \sqrt{\pi  C+S} \sqrt{4 \pi ^3 C J^2+\pi  C S^2+2 \pi ^2 Q^2 S+S^3}}{-4 \pi ^4 C^3 J^2+4 \pi  C^2 S^3+\pi ^2 C^3 S^2+2 \pi ^2 C Q^2 S^2+3 C S^4}
	\end{equation} 
	The entropy is plotted against $\tau$ at $C=0.2,Q=1.5,J=0.02$ and $l=1$ in Figure.\ref{1}. Here we observe three black hole branches: a small, an intermediate, and a large black hole branch. We also see a generation point at  $\tau=1.84076, S=2.02107$ and an annihilation point at $\tau=1.27059, S=  0.07658$ which are shown as blue dot and black dot respectively.\\
	
	We now calculate the topological charge of the black hole using the residue method.	The complex function $\mathcal{R}(z)$ mentioned in section \textbf{I} is given by : 	
	\begin{equation}
	\mathcal{R}(z)=-\frac{\alpha_1}{\beta_1}
	\label{res}
	\end{equation}
	where 
	\begin{equation*}
	\alpha_1=-4 \pi ^4 C^2 J^2+\pi ^2 C^2 z^2+4 \pi  C z^3+2 \pi ^2 Q^2 z^2+3 z^4
	\end{equation*}
	and the denominator of the complex function is considered as a polynomial function $\mathcal{A}(z)$ 
	\begin{multline}
	\mathcal{A}(z)=\beta_1=4 \pi ^4 C^2 J^2 \tau -\pi ^2 C^2 \tau  z^2+4 \pi ^{3/2} l z \sqrt{C z} \sqrt{\pi  C+z} \sqrt{4 \pi ^3 C J^2+\pi  C z^2+2 \pi ^2 Q^2 z+z^3}\\-4 \pi  C \tau  z^3-2 \pi ^2 Q^2 \tau  z^2-3 \tau  z^4
	\label{pole}
	\end{multline}
	
	To calculate the topological charge, we put $C=0.2,Q=1.5,J=0.02$ and $l=1$ in equation \ref{pole}.
	\begin{figure}[h]
		\centering
		\includegraphics[height=6cm,width=10cm]{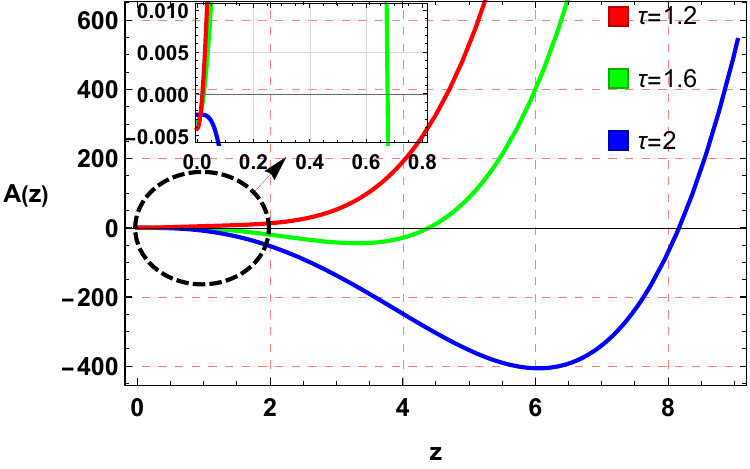}
		\caption{Plot for polynomial function $\mathcal{A}(z)$ for $\tau=2.2,2.4$ and $2.6$}
		\label{2}
	\end{figure}	

\begin{figure}[h]	
	\centering
	\begin{subfigure}{0.42\textwidth}
		\includegraphics[width=\linewidth]{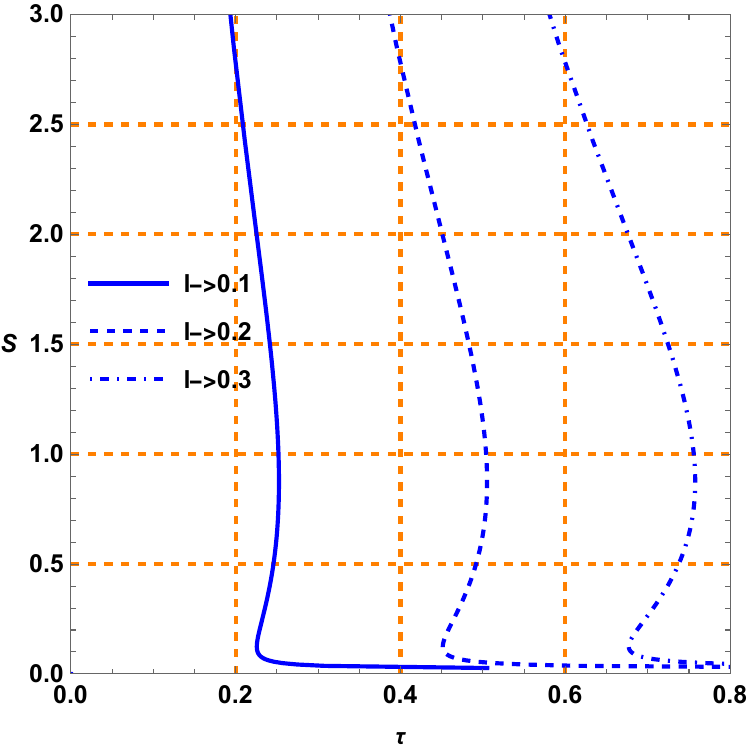}
		\caption{W=1}
		\label{3a}
	\end{subfigure}
	\hspace{1cm}
	\begin{subfigure}{0.4\textwidth}
		\includegraphics[width=\linewidth]{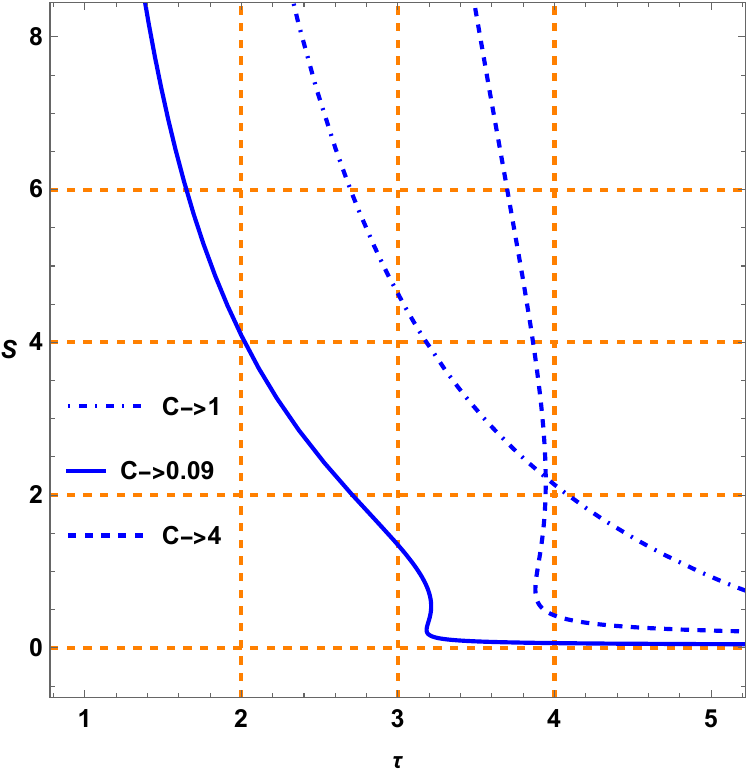}
		\caption{W=1}
		\label{3b}
	\end{subfigure}
	\begin{subfigure}{0.41\textwidth}
		\includegraphics[width=\linewidth]{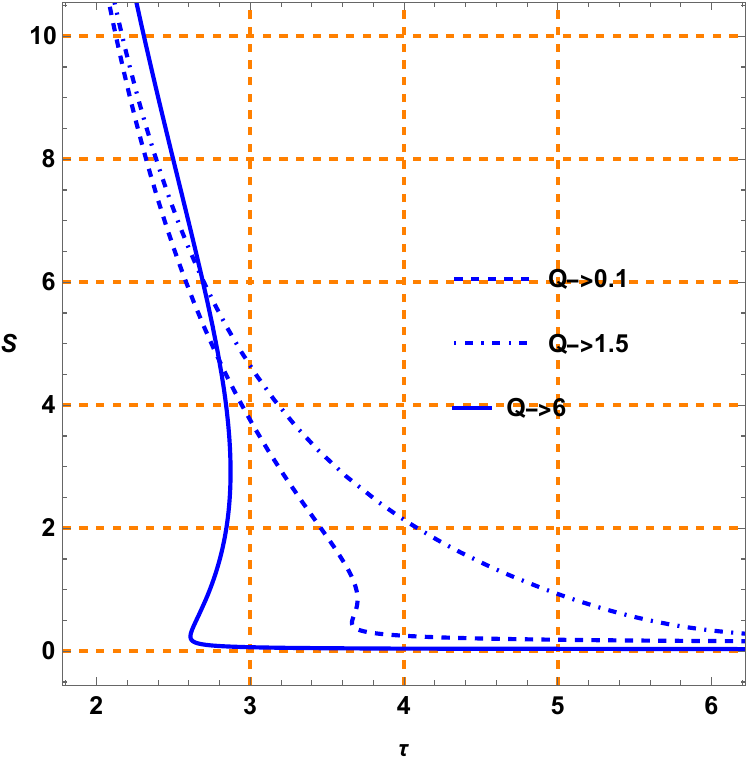}
		\caption{W=1}
		\label{3c}
	\end{subfigure}
	\hspace{1cm}
	\begin{subfigure}{0.4\textwidth}
			\includegraphics[width=\linewidth]{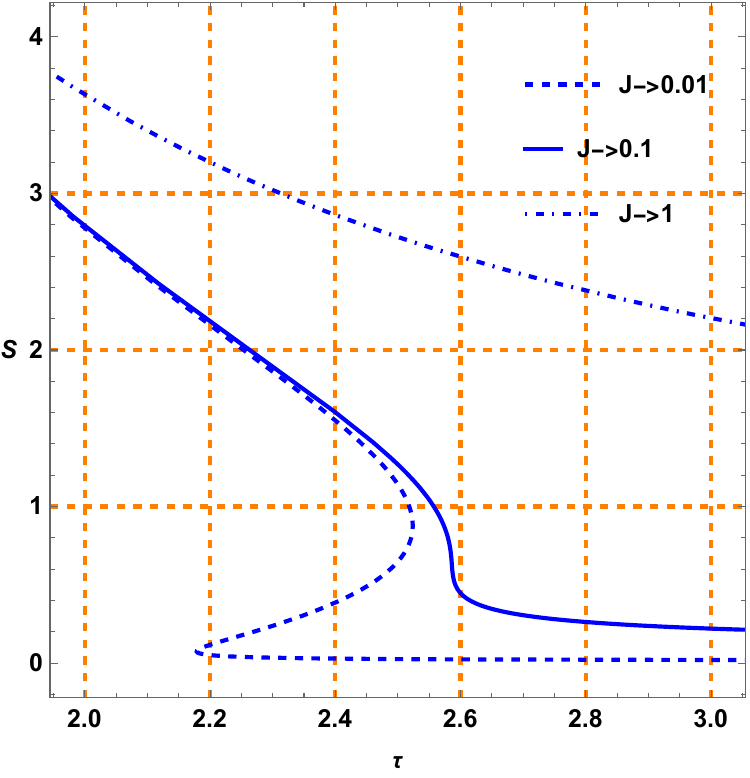}
			\caption{W=1}
			\label{3d}
		\end{subfigure}
	\caption{ $\tau$ vs $S$ plots for Kerr-Sen-Ads black hole in restricted phase space in fixed $(q,J,C)$ ensemble. Here $\tau$ is plotted against $S$ for  different $l$ values while keeping $C=0.2,Q=1.5,J=0.02$ fixed. Fig (b) represents the variation of $\tau$ vs $S$ plots for different $C$ values while keeping $l=1,Q=1.5,J=0.02$ fixed. Fig(c) shows variation of $\tau$ vs $S$ plots for different $Q$ values while keeping $l=1,C=1,J=0.02$ fixed and finally Fig.(d) shows the variation of $\tau$ vs $S$ plots for different $Q$ values while maintaining  constant $l,Q,J$ at $l=1,Q=1.5,C=0.2$. $W$ denotes topological charge.}
	\label{3}
\end{figure}

Figure \ref{2} represents the plot for $\mathcal{A}(z)$. This demonstrates that there is one pole for $\tau<1.27059$, three poles for $1.27059 < \tau < 1.84075 $, and one pole for $\tau > 1.84075$. Using $\tau=1.2$ for the large black hole branch, the pole is located at $z=8.16619$ as suggested by the blue solid line in figure \ref{2}. The winding number is $w=+1$ around the pole $z=8.16619$. Using $\tau=1.6$, the poles for the intermediate black hole branch are located at $z_1=0.0235863,z_2=0.676384$, and $z_3=4.39265$, which is represented by the solid green line. The winding numbers around these three poles are $w_1=+1, w_2=-1$, and $w_3=+1$, respectively, based on the sign of the residue around them. Thus, $$W=w_1+w_2+w_3=1-1+1=1$$ is the topological charge. Similarly, as represented by the solid red line, for a small black hole branch, taking $\tau=2$, the pole is located at $z=0.0186124$, and the winding number is determined to be $+1$ based on the sign of residue surrounding the pole.\\
We now analyze the effect of thermodynamic parameters on the topological charge. We start with analysing the effect of Ads radius $l$ keeping $C=0.2,Q=1.5,J=0.02$ constant. We observe no effect of change in $l$ on the topological charge. The same is illustrated in figure \ref{3a}. Here we have considered three values of $l$ keeping  $C=0.2,Q=1.5,J=0.02$ constant. For all the cases, we find the number of black hole branches equal to three and the topological charge to be one.\\

Our above analysis is followed by studying the dependence of topological charge on the central charge $C$ keeping $l=1,Q=1.5,J=0.02$ fixed. It is seen that topological charge does not depend on variation in central charge $C$. Although for a specific range of $C$ the three phases of the black hole merge into a single branch. However it does not alter the topological number. In figure \ref{3b}, for $C=0.09$ and $C=4$ three black hole branches and topological charge $1$ are observed. On the other hand for $C=1$, we have only a single black hole branch but the topological charge is still found to be $1.$  \\

Next, we want to vary $Q$ while holding $l=1,C=1,J=0.02$ constant in order to see how the change in charge affects the topological charge. In Figures \ref{3c},we have set $Q=0.1$, $Q=1.5$, and $Q=6$. Three black hole branches are found for $Q=0.1$ and $Q=6$ while one black hole branch is found for $Q=1.5$.In each of the three scenarios, the topological charge is $1$. Therefore, it is noted that charge likewise has no effect on the topological number.\\

	Lastly, we perform the same study in figure \ref{3d}, to examine how $J$ affects topological charge while maintaining constant $l,Q,J$ at $l=1,Q=1.5,C=0.2$. J is set to 0.02, 0.1 and 1 in figure \ref{3c}.Topological charge in each of the three scenarios is $1$. Thus, it is concluded that while the number of black hole branches may change in response to changes in angular momentum $J$, the topological charge remains constant at $W=1$.

In conclusion, our findings show that the Kerr-Sen-Ads black hole in a fixed $(Q,J,C)$ ensemble, the topological charge is $1$. The topological charge is independent of the thermodynamic parameters $Q,J,C$.\\
	\subsection{Fixed ($\phi$,J,C) Ensemble}
\begin{figure}[h!t]
	\centering
	\includegraphics[width=10cm,height=8cm]{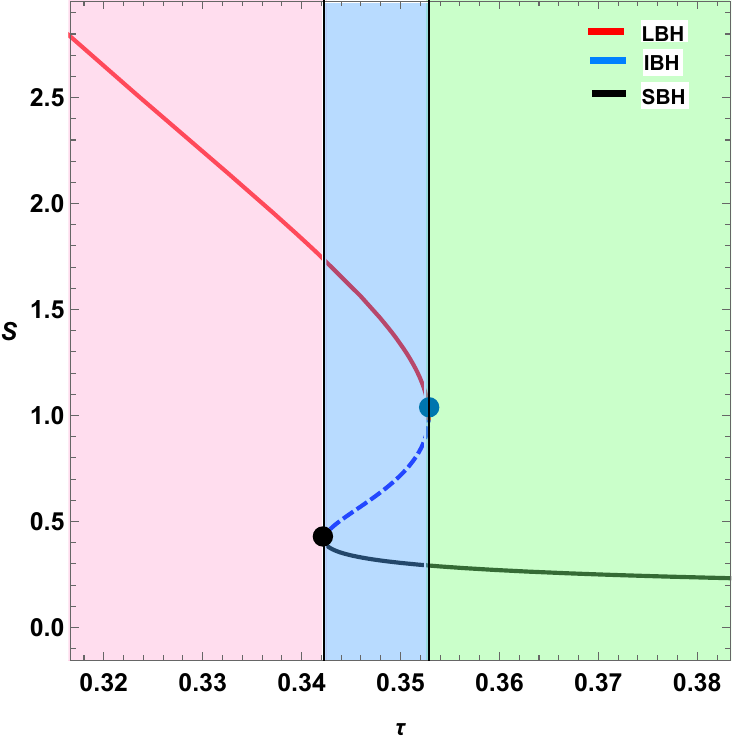}
	\caption{$\tau$ vs $S$ plot for Kerr Sen Ads  black hole at  $C=1,\phi=1.5,J=0.02$ and $l=0.1$. The $LBH$ and $SBH$ regions represent the large and small black hole branches respectively.}
	\label{4}
\end{figure}

\begin{figure}[h]	
	\centering
	\begin{subfigure}{0.3\textwidth}
		\includegraphics[width=\linewidth]{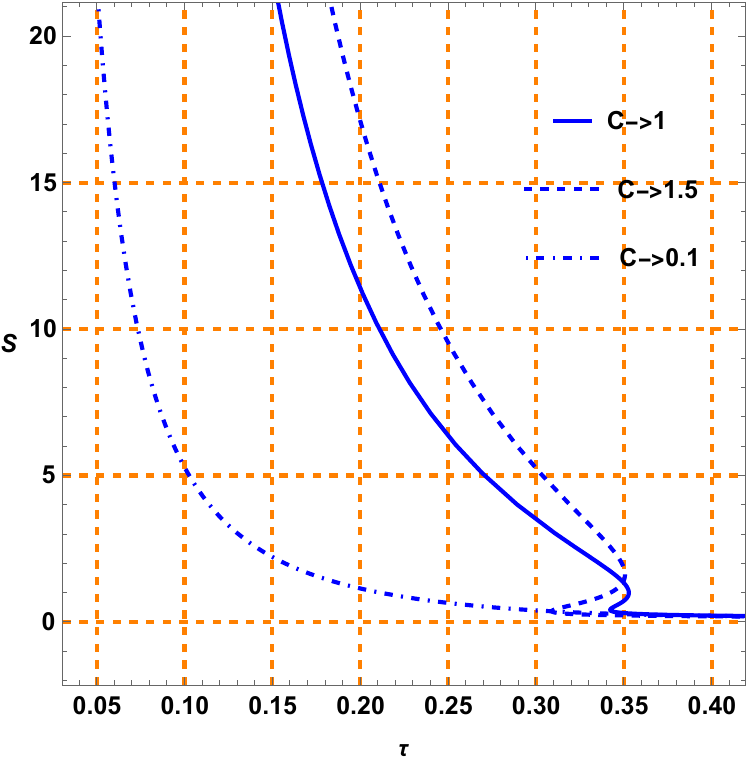}
		\caption{W=1}
		\label{5a}
	\end{subfigure}
	\begin{subfigure}{0.3\textwidth}
		\includegraphics[width=\linewidth]{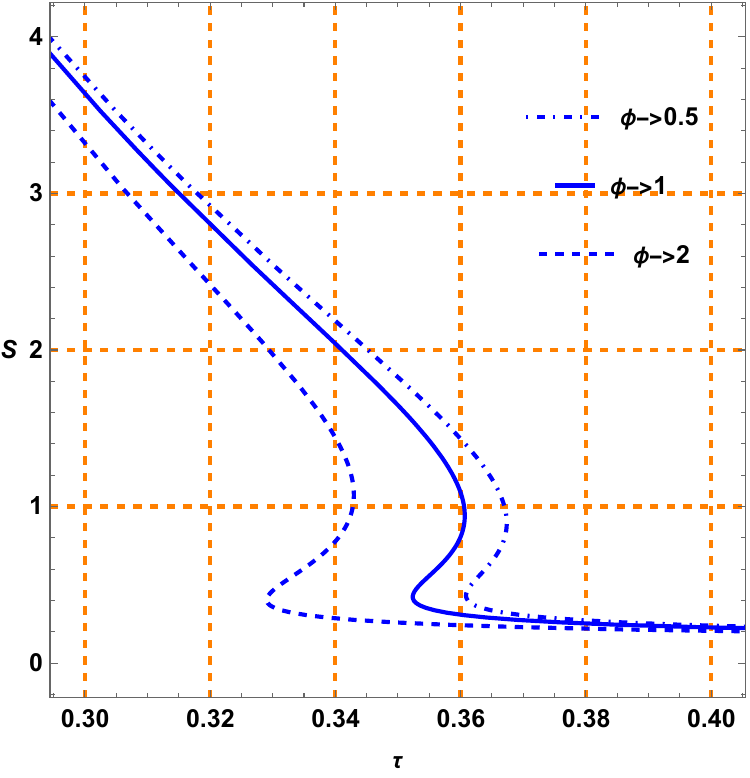}
		\caption{W=1}
		\label{5b}
	\end{subfigure}
	\begin{subfigure}{0.32\textwidth}
		\includegraphics[width=\linewidth]{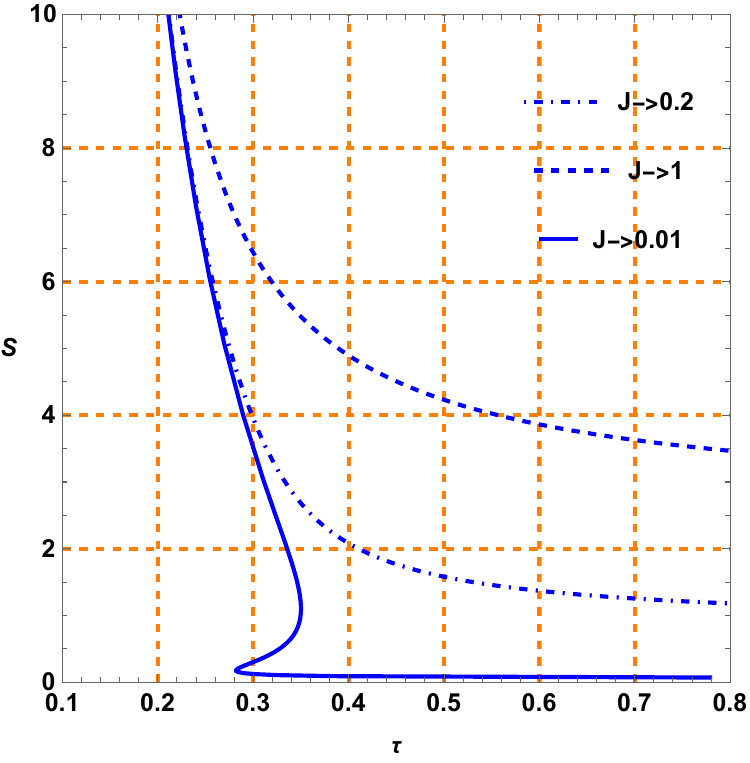}
		\caption{W=1}
		\label{5c}
	\end{subfigure}
	\caption{ Fig (a) is the variation of $\tau$ vs $S$ plots with $C$ for Kerr-Sen-Ads black hole in restricted phase space in fixed $(\phi,J,C)$ ensemble. Here $\tau$ is plotted against $S$ for different $C$ values while keeping $l=0.1,\phi=1.5,J=0.02$ fixed.  Fig (b) illustrates the variation of $\tau$ vs $S$ plots with $\phi$ . Here $\tau$ is plotted against $S$ for different $\phi$ values while keeping $l=0.1,C=1,J=0.02$ fixed.Fig (c) represents the variation of $\tau$ vs $S$ plot with $J$ while keeping $l=1,C=0.2,\phi=1.5$ fixed. $W$ denotes the topological charge.
	}
	\label{5}
\end{figure}

	The potential $\phi$, angular momentum $J$ and central charge $C$ is kept fixed in fixed ($\phi$,J,C) ensemble. The potential $\phi$ is given by :
\begin{equation}
	\phi=\frac{\partial M}{\partial q}=\frac{\sqrt{\pi } Q S \sqrt{\pi  C+S}}{l \sqrt{C S} \sqrt{\pi  C \left(4 \pi ^2 J^2+S^2\right)+2 \pi ^2 Q^2 S+S^3}}
	\label{phi2}
\end{equation}
To get an expression for $q$ we solve equation \ref{phi2} and obtain :
\begin{equation}
	Q=\frac{\sqrt{\pi } \sqrt{C} l \phi  \sqrt{4 \pi ^3 C J^2+\pi  C S^2+S^3}}{\sqrt{S} \sqrt{-2 \pi ^2 C l^2 \phi ^2+\pi  C+S}}
	\label{q}
\end{equation}
Now the new mass($M_{\phi}$) in this ensemble is calculated using equation \ref{q} as follows :
\begin{multline}
	M_{\phi}=M-q \phi\\=\frac{\sqrt{\pi  C+S} \sqrt{\frac{2 \pi ^3 C l^2 \phi ^2 \left(4 \pi ^3 C J^2+\pi  C S^2+S^3\right)}{-2 \pi ^2 C l^2 \phi ^2+\pi  C+S}+\pi  C \left(4 \pi ^2 J^2+S^2\right)+S^3}}{2 \pi ^{3/2} l \sqrt{C S}}
	-\frac{\sqrt{\pi } \sqrt{C} l \phi ^2 \sqrt{4 \pi ^3 C J^2+\pi  C S^2+S^3}}{\sqrt{S} \sqrt{-2 \pi ^2 C l^2 \phi ^2+\pi  C+S}}
\end{multline}
The off-shell free energy is computed using:
$$\mathcal{F}=M_{\phi}-S/\tau$$
We compute the expressions for $\phi^S$ and $\tau$ using the same method as in the preceding part.\\
In Figure.\ref{4}, the entropy $S$ vs $\tau$ is plotted at $C=1,\phi=1.5,J=0.02$, and $l=0.1$. Here also three black hole branches are visible. We see two points that are: a generation point at $\tau=0.35285, S= 0.9876$ and an annihilation point at $\tau=0.342307, S=0.4037$, shown as blue and black dots, respectively\\
To analyze the effect of thermodynamic parameters on the topological charge we vary one particular parameter keeping other parameters fixed. For this ensemble, we observe no effect of change in any thermodynamic parameter on the topological charge. The same is illustrated in figure \ref{5} \\

In figure \ref{5a}, central charge $C$ is varied keeping $l=0.1,\phi=1.5,J=0.02$ fixed. For a specific range of $C$ the three phases of the black hole are visible and it merges into a single branch for other values, throughout the process the topological charge remains constant at $1.$ \\

In figure \ref{5b}, the effect of variation of $\phi$ in the topological charge is shown while keeping $l=0.1,C=1,J=0.02$ constant. We have set $\phi=0.5$, $\phi=1$, and $\phi=2$. Three black hole branches are found in each of the three scenarios and the topological charge is still computed to be $1$. Therefore, it is noted that potential $\phi$ has no effect on the topological number.\\

Lastly,  figure \ref{5c} is plotted to check how $J$ affects topological charge while keeping constant $l,\phi,C$ at $l=1,\phi=1.5,C=1$. In all three cases presented in Fig.\ref{5c}, the topological charge is $1$. Consequently, it is inferred that the topological charge at $W=1$ remains constant, however, the number of black hole branches may vary in response to variations in angular momentum $J$.\\
In conclusion, our findings show that the Kerr-Sen-Ads black hole in a fixed $(Q,J,C)$ ensemble, the topological charge is $1$. The topological charge is independent of the thermodynamic parameters $Q,J,C$.\\
\subsection{Fixed (Q,$\Omega$,C) Ensemble}
\begin{figure}[h!t]
	\centering
	\includegraphics[width=10cm,height=9cm]{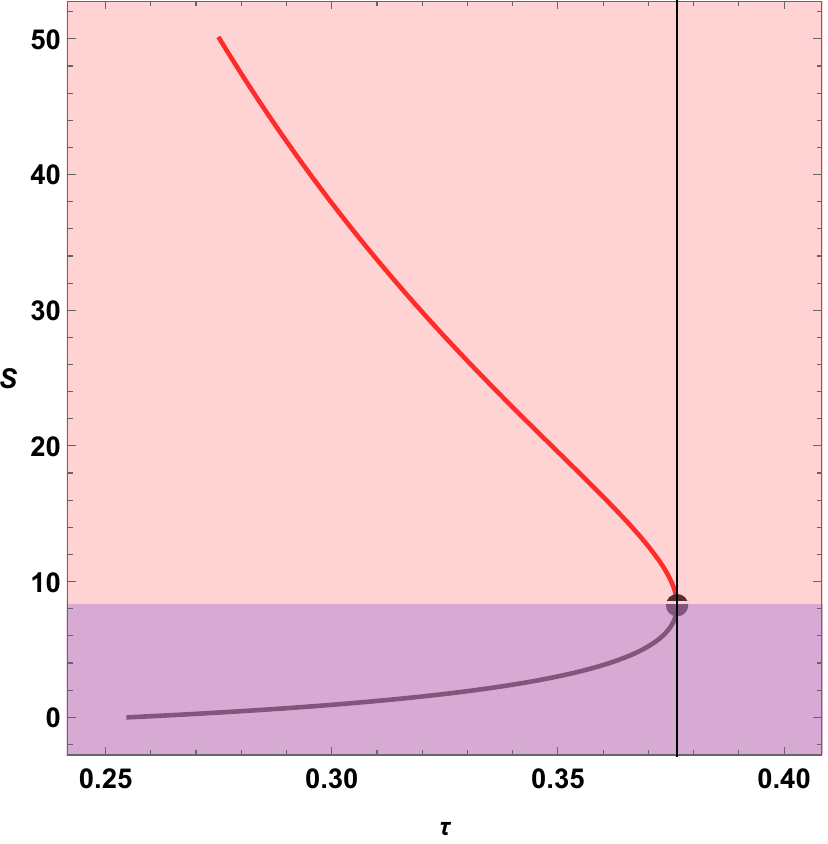}
	\caption{$\tau$ vs $S$ plot for Kerr Sen Ads  black hole at  $C=1,\phi=1.5,J=0.02$ and $l=0.1$. The red and the black coloured solid represent the large and small black hole branches respectively.}
	\label{11}
\end{figure}

\begin{figure}[h!t]
	\centering
		\begin{subfigure}{0.4\textwidth}
		\includegraphics[width=\linewidth]{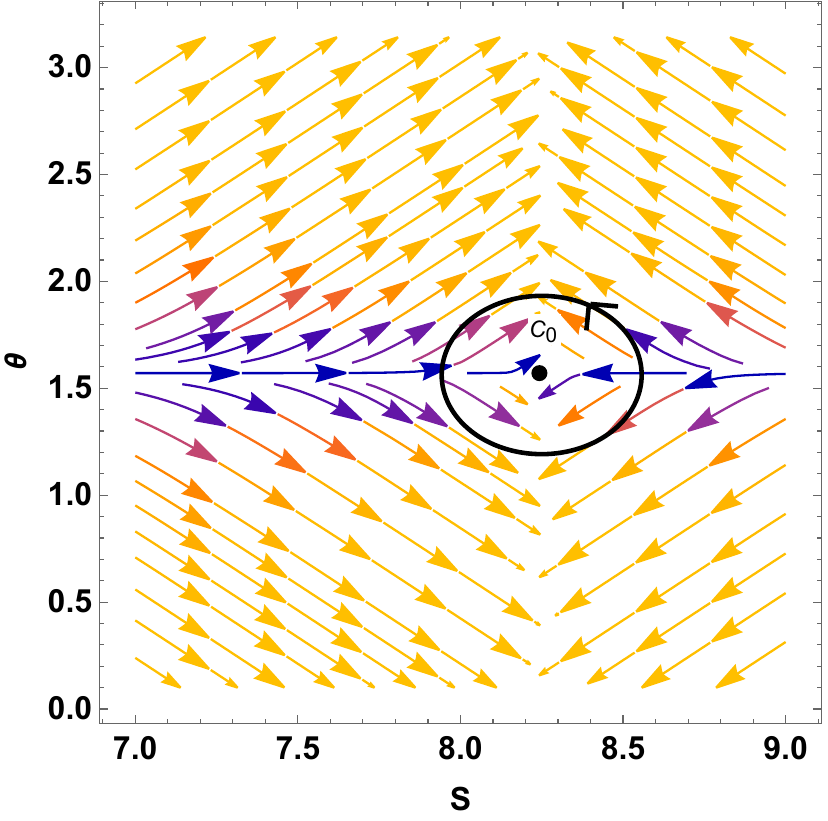}
		\caption{}
		\label{12a}
	\end{subfigure}
	\begin{subfigure}{0.4\textwidth}
	\includegraphics[width=\linewidth]{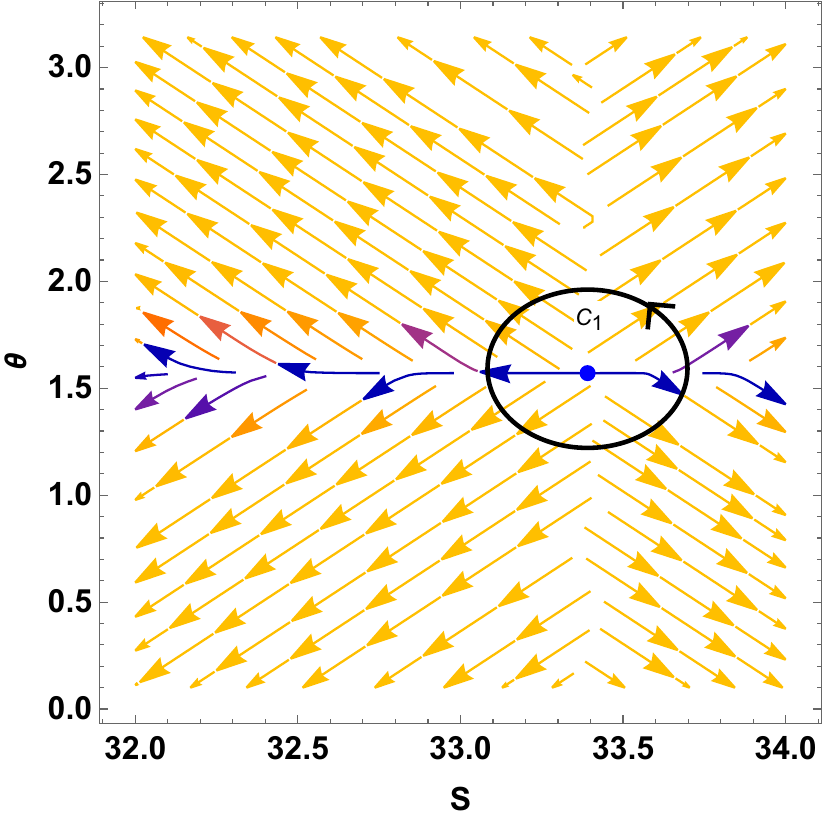}
	\caption{}
	\label{12b}
\end{subfigure}
	\caption{A portion of the vector field $\phi$ on $S-\theta$ plane. The black dot of Figure $(a)$ represents Davies point and the blue dot in Figure $(b)$ represents Hawking page phase transition point.}
	\label{12}
\end{figure}

\begin{figure}[h]	
 	\centering
 	\begin{subfigure}{0.40\textwidth}
 		\includegraphics[width=\linewidth]{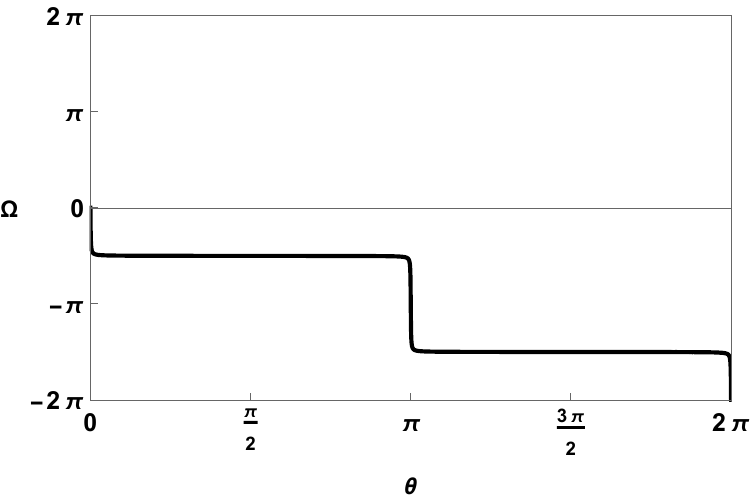}
 		\caption{}
 		\label{12.1a}
 	\end{subfigure}
 	\hspace{0.5cm}
 	\begin{subfigure}{0.40\textwidth}
 		\includegraphics[width=\linewidth]{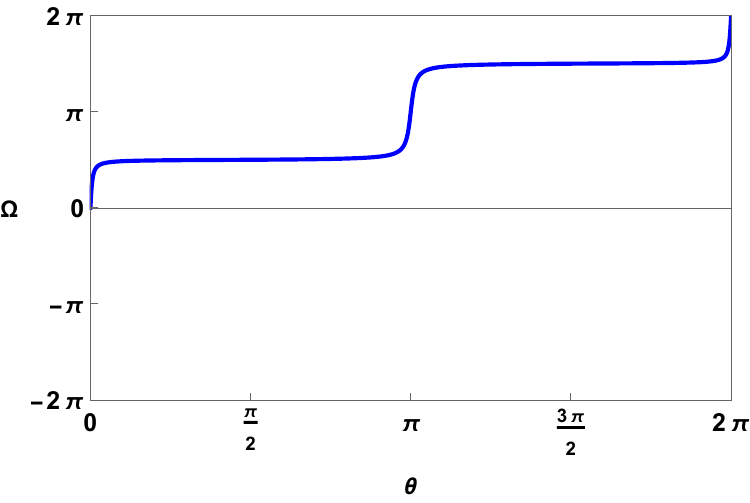}
 		\caption{}
 		\label{12.2b}
 	\end{subfigure}
 	
 	\caption{Kerr Sen AdS black hole in fixed ($\phi$,J,C) ensemble: Figure (a) represents topological number calculation  for Davies point around the $C_0$ contour in Figure.\ref{12a} and Figure (b) represents the topological number calculation  for Hawking-Page phase transition point around the $C_1$ contour
 	}
 	\label{12.1}
 \end{figure}

\begin{figure}[h]	
	\centering
	\begin{subfigure}{0.3\textwidth}
		\includegraphics[width=\linewidth]{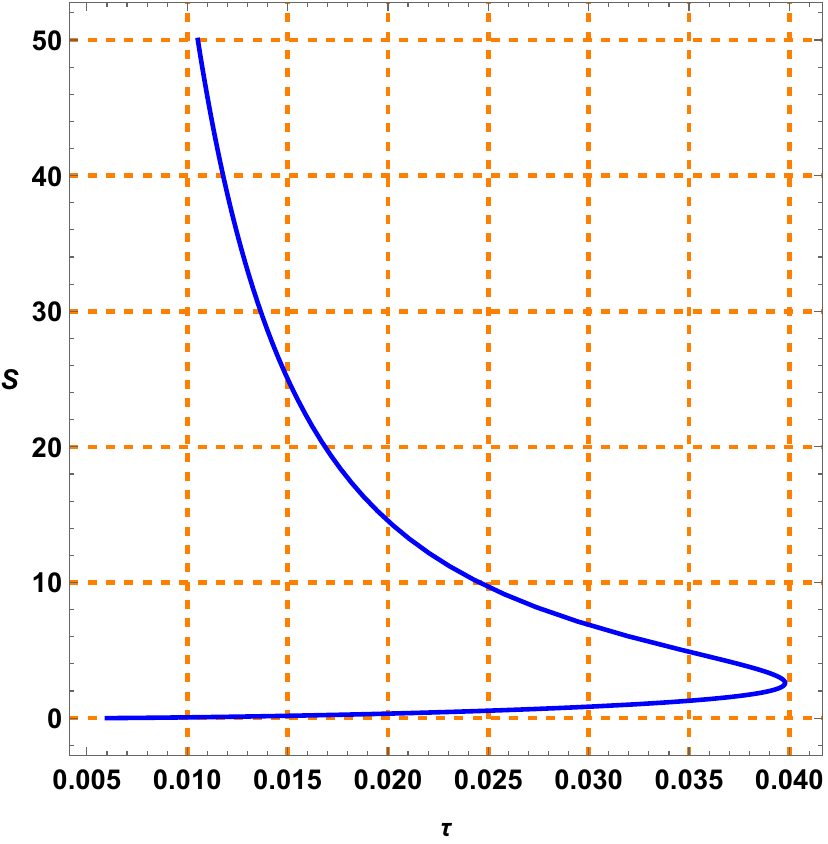}
		\caption{W=0}
		\label{13a}
	\end{subfigure}
	\begin{subfigure}{0.3\textwidth}
		\includegraphics[width=\linewidth]{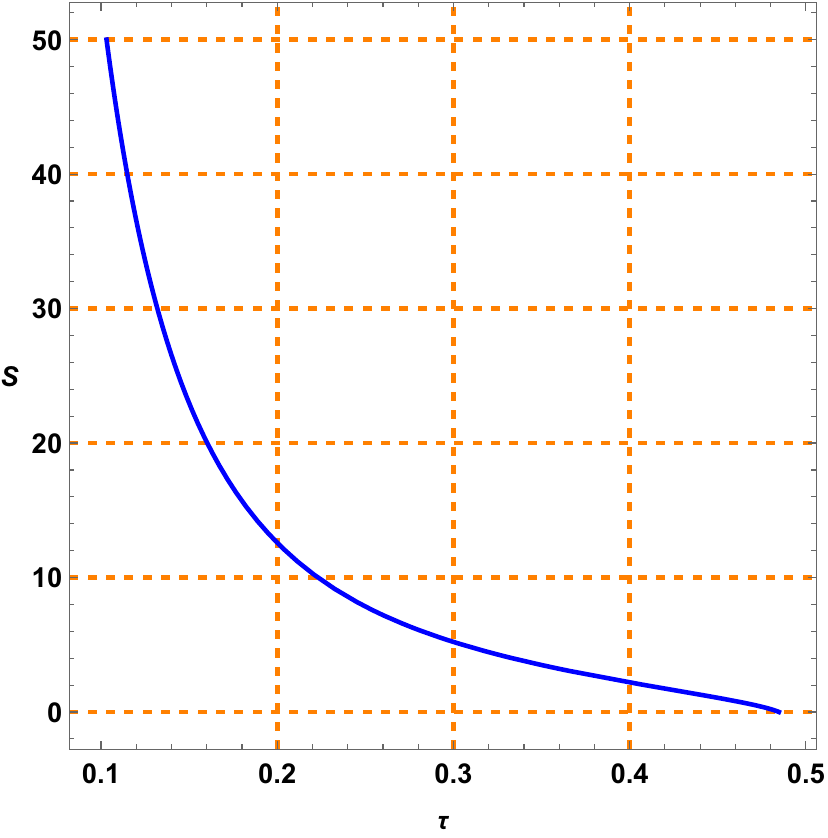}
		\caption{W=1}
		\label{13b}
	\end{subfigure}
	\begin{subfigure}{0.3\textwidth}
	\includegraphics[width=\linewidth]{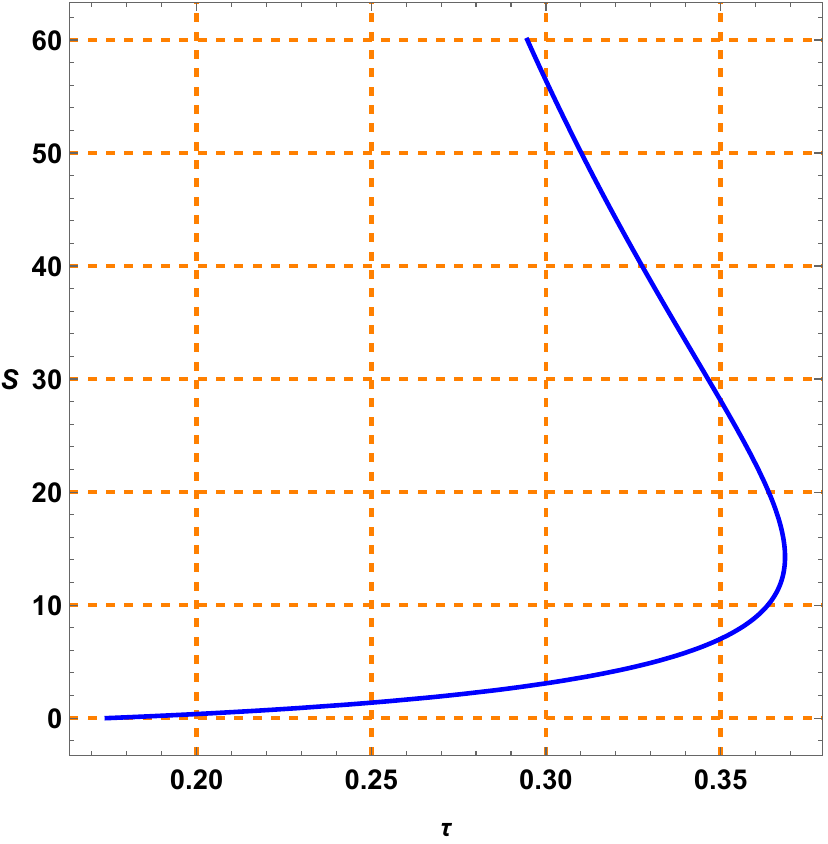}
	\caption{W=0}
	\label{13c}
\end{subfigure}
	\caption{ $\tau$ vs $S$ plots for Kerr-Sen-Ads black hole in restricted phase space in fixed $(Q,\Omega,C)$ ensemble. Here $\tau$ is plotted against $S$ for two different $C$ values while keeping $l=0.1,Q=1.5,\Omega=0.02$ fixed. Figure $\left(a\right)$ shows $\tau$ vs $S$ plot at $C=0.01$, figure $\left(b\right)$ shows $\tau$ vs $S$ plot  at $C=0.1$, figure $\left(c\right)$ shows the same at $C=15$ $W$ denotes topological charge.
	}
	\label{13}
\end{figure}

In this ensemble, the angular frequency $\Omega$, charge $Q$ and central charge $C$ are kept fixed. The angular frequency $\Omega$ is given by :
\begin{equation}
	\Omega=\frac{2 \pi ^{3/2} C J \sqrt{\pi  C+S}}{l \sqrt{C S} \sqrt{\pi  C \left(4 \pi ^2 J^2+S^2\right)+2 \pi ^2 Q^2 S+S^3}}
	\label{omega3}
\end{equation}
By solving equation \ref{omega3}, we get the expression for $J$ as :
\begin{equation}
	J=\frac{l S \Omega  \sqrt{\pi  C S+2 \pi ^2 Q^2+S^2}}{\pi ^{3/2} \sqrt{C} \sqrt{4 \pi  C-4 l^2 S \Omega ^2+4 S}}
	\label{j}
\end{equation}
The newly mass($M_{\Omega}$) in this ensemble :
\begin{multline}
	M_{\Omega}=M-J \Omega\\=-\frac{l^2 S \Omega ^2 \sqrt{C S} \sqrt{\pi  C S+2 \pi ^2 Q^2+S^2}-\sqrt{C} \sqrt{\pi  C+S} \sqrt{\pi  C-l^2 S \Omega ^2+S} \sqrt{\frac{S (\pi  C+S) \left(\pi  C S+2 \pi ^2 Q^2+S^2\right)}{\pi  C-l^2 S \Omega ^2+S}}}{2 \pi ^{3/2} \sqrt{C} l \sqrt{C S} \sqrt{\pi  C-l^2 S \Omega ^2+S}}
\end{multline}
The off-shell free energy is computed using:
$$\mathcal{F}=M_{\Omega}-S/\tau$$
We compute the expressions for $\phi^S$ and $\tau$ using the same method as in the preceding part.\\
In Figure.\ref{11}, the entropy $S$ vs $\tau$ is plotted at $C=10,Q=1.5,\Omega=0.02$, and $l=0.1$. We observe two black hole branches with a generation point at $(\tau,S)$ = $(0.37635,8.2427)$ represented by the black dot. We observe a large black hole branch shown as the red solid line and a small black hole branch shown as the black solid line. The topological number of the smaller black hole branch is $-1$ and that of the large black hole branch is $+1$. The total topological number is found to be $-1+1=0.$\\
Taking a cue from figure.\ref{11}, the study of Hawking-Page transition and Davies type of phase transition also can be studied by constructing a novel vector field given by \cite{newfield}
\begin{equation}
	\phi=\left(\frac{1}{S}\frac{\partial F^2}{\partial S},-cot\theta csc\theta\right)
	\label{newf}
\end{equation}
In equation eq.\ref{newf}, a new two-dimensional vector field is introduced to facilitate the analysis of the topology associated with the Hawking-Page transition and the Davies-type phase transition, as demonstrated in \cite{newfield}. This vector field is constructed in such a way that its critical points, or zero points, coincide with the points where the Hawking-Page transition and Davies point occur. 
\textbf{In equation  eq.\ref{newf},   \( F = M - TS \) represents the free energy, and \( T = \frac{\partial M}{\partial S} \) is the temperature. By design, the condition \(\phi^S = 0\) leads to:}
\begin{equation}
    F = 0 \hspace{0.5 cm} \text{and} \hspace{0.5 cm} \frac{1}{C} = 0
\end{equation}
Here, \( F = 0 \) signifies the Hawking-Page phase transition, while \( \frac{1}{C} = 0 \) marks the Davies point. Thus, the zero points of this vector field align precisely with the Hawking-Page and Davies phase transition points, confirming the correspondence between these critical points and the transitions in the thermodynamic phase structure.
From the vector plot shown in the figure.\ref{12}
By solving $F=0$ and $C_\Omega=T\frac{\partial S}{\partial T}=0$, the Hawking Page phase transition point is found to be $S=33.3908$ and the Davies point if found to be $S=8.2427$ which exactly matches with the critical points of the vector field in equation \ref{new}. The black dot ($S= 8.2427$) in figure \ref{12a} is a Davies point and the blue dot ($S=33.3908$) is a Hawking page phase transition point. \\
The topological charge can be calculated by constructing a contour $C$ around each zero point($C_0$ and $C_1$ as shown in figure \ref{12}) which is parameterized as:
\begin{equation}
	\begin{cases}
		S=a cos\nu+r_0\\
		\theta=b sin \nu +\frac{\pi}{2}\\
	\end{cases}
\end{equation}
where $\nu \in (0,2\pi)$. To calculate the deflection $\Omega$ of the vector field $n$ along the contour $C_1$ and $C_2$, we choose $a=b=0.2$ and $r_0= \text{the zero point itself enclosed by the contours}.$ We use the formula \cite{g1,g2}:
\begin{equation}
	\Omega(\nu)=\int_{0}^{\nu} \epsilon_{12} n^1\partial_\nu n^2 d\nu
\end{equation}
where the unit vectors $\left(n^1, n^2\right)$ are given by: 
$$n^1=\frac{\phi^S}{\sqrt{(\phi^S)^2+(\phi^\Theta)^2}}  \hspace{0.5cm}\text{and} \hspace{0.5cm} n^2=\frac{\phi^\theta}{\sqrt{(\phi^S)^2+(\phi^\Theta)^2}}$$
Finally, the winding numbers $w$ and topological charge $W$ can be calculated as follows:
\begin{equation}
	\begin{cases}
		w=\frac{1}{2\pi} \Omega(2\pi)\\
		W=\sum_{i} w_i\\
	\end{cases}
\end{equation}
In cases where the parameter region does not encompass any zero points, the topological charge equals to $0$. Following the above procedure, we plot deflection $\Omega$ vs $\theta$ plot in figure.\ref{12.1}. From the figure, the topological charge of the Hawking page phase transition(blue-coloured solid line) is $+1$ and that of Davis point(black-coloured solid line) is found to be $-1.$ \\
To analyze the effect of thermodynamic parameters on the topological charge we vary one particular parameter keeping other parameters fixed.
In figure \ref{13}, the value of central charge $C$ is varied keeping $l=0.1,Q=1.5,\Omega=0.02$ fixed. For a different value of $C$ the topological charge is found to be $0$ or $1.$ For $C=0.01$, in figure \ref{13a} there are two black hole branches and topological charge is $0$. In \ref{13b}, we have only a single black hole branch with a topological charge of $1$ when $C=1$. In \ref{13c} again we observe two black hole branches with a topological charge of $0$ when $C=15$. \\

\begin{figure}[h]	
	\centering
	\begin{subfigure}{0.3\textwidth}
		\includegraphics[width=\linewidth]{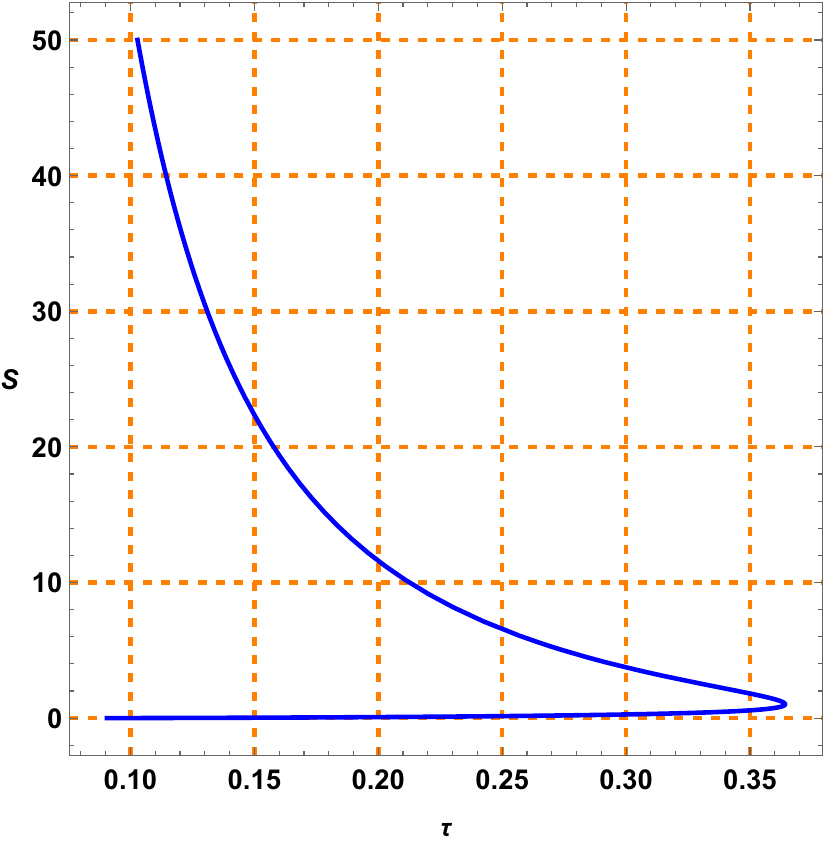}
		\caption{W=0}
		\label{14a}
	\end{subfigure}
	\begin{subfigure}{0.3\textwidth}
		\includegraphics[width=\linewidth]{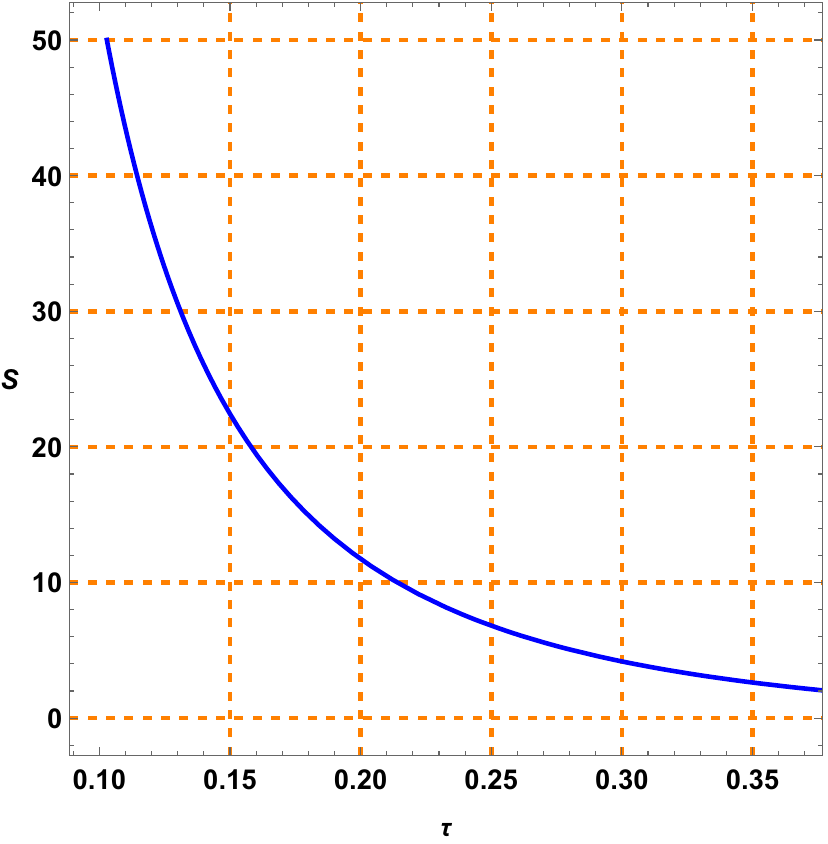}
		\caption{W=1}
		\label{14b}
	\end{subfigure}
	\begin{subfigure}{0.3\textwidth}
		\includegraphics[width=\linewidth]{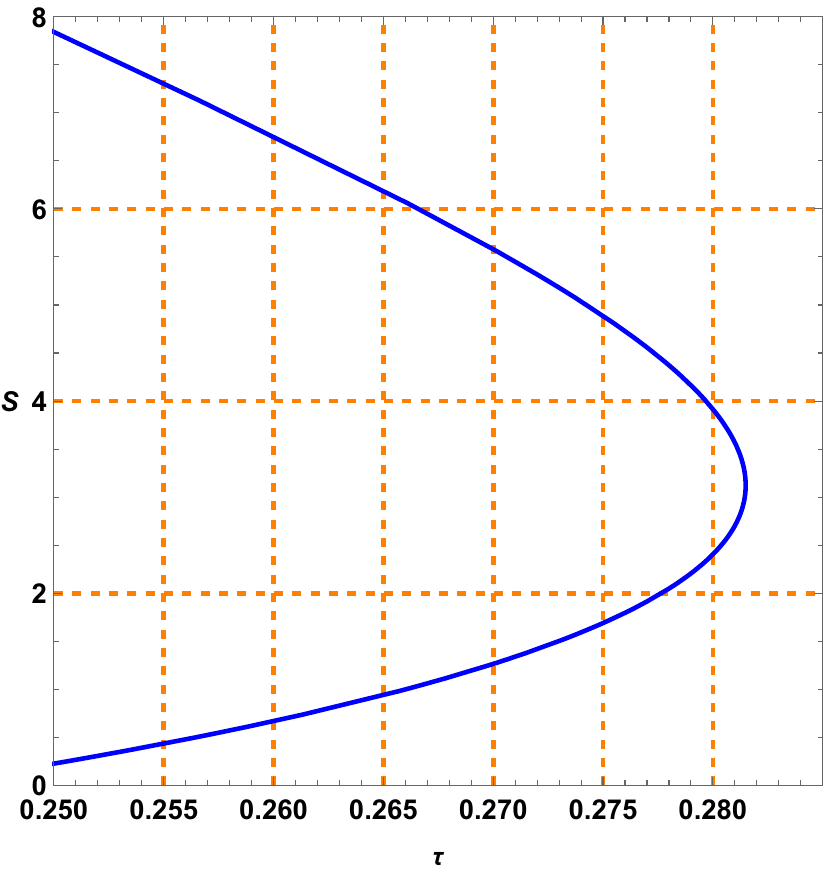}
		\caption{W=0}
		\label{14c}
	\end{subfigure}
	
	\caption{ $\tau$ vs $S$ plots for Kerr-Sen-Ads black hole in restricted phase space in fixed $(\phi,J,C)$ ensemble. Here $\tau$ is plotted against $S$ for three different $\phi$ values while keeping $l=0.1,C=1,J=0.02$ fixed. Figure $\left(a\right)$ shows $\tau$ vs $S$ plot at $\phi=0.5$,figure $\left(b\right)$ shows $\tau$ vs $S$ plot  at $\phi=1$ and in figure $\left(c\right)$ shows $\tau$ vs $S$ plot at $\phi=5$.$W$ denotes topological charge.
	}
	\label{14}
\end{figure}

\begin{figure}[h]	
	\centering
	\begin{subfigure}{0.3\textwidth}
		\includegraphics[width=\linewidth]{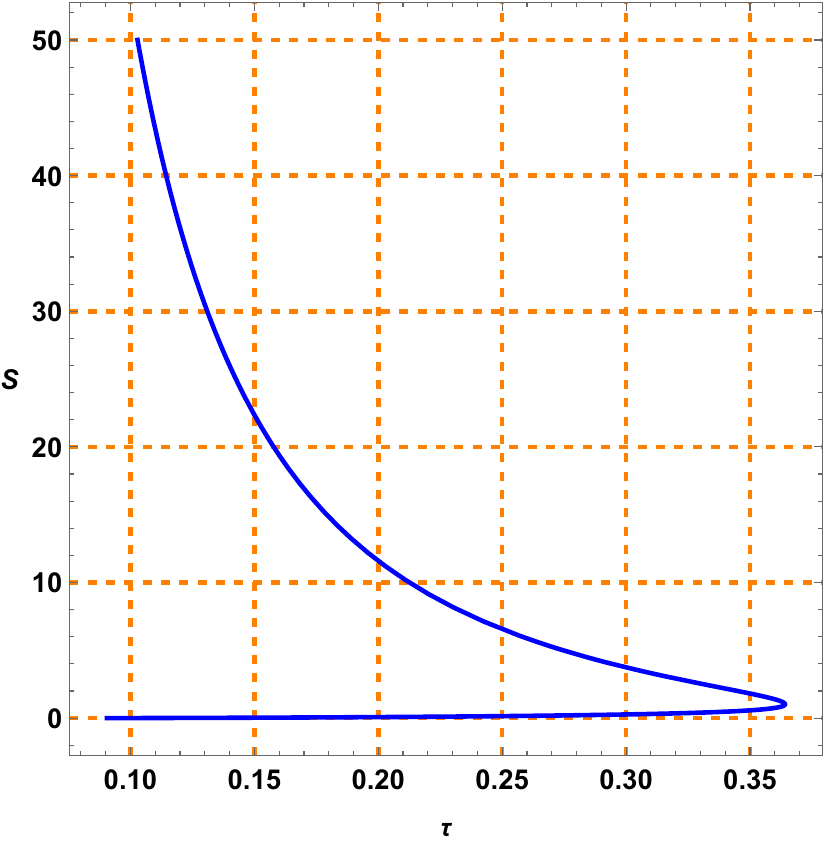}
		\caption{W=0}
		\label{15a}
	\end{subfigure}
	\begin{subfigure}{0.3\textwidth}
		\includegraphics[width=\linewidth]{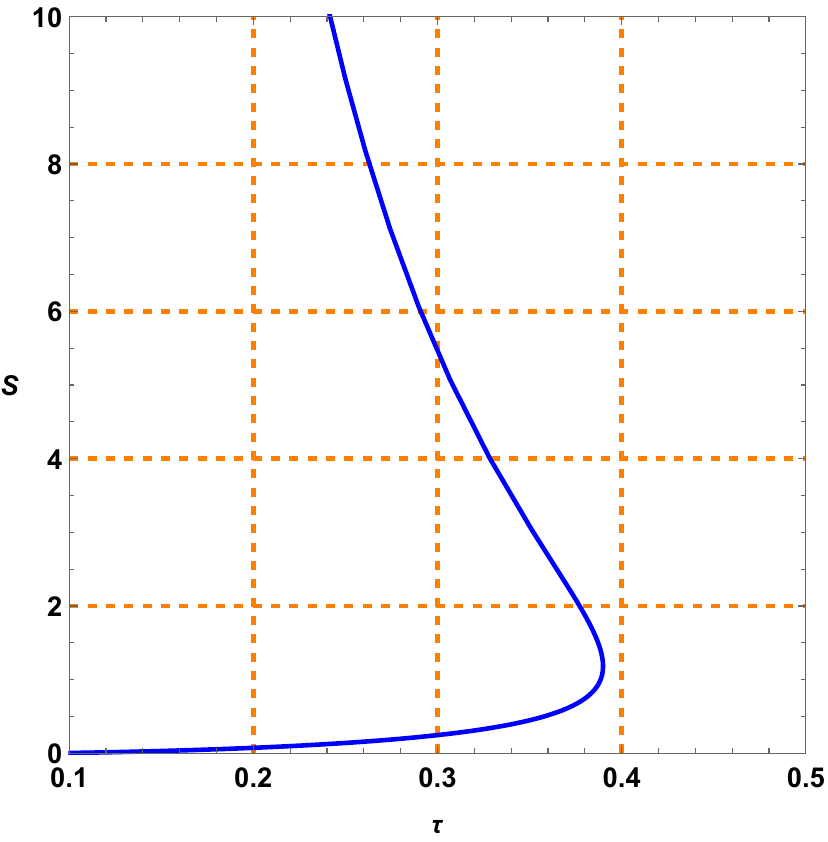}
		\caption{W=0}
		\label{15b}
	\end{subfigure}
	\begin{subfigure}{0.3\textwidth}
		\includegraphics[width=\linewidth]{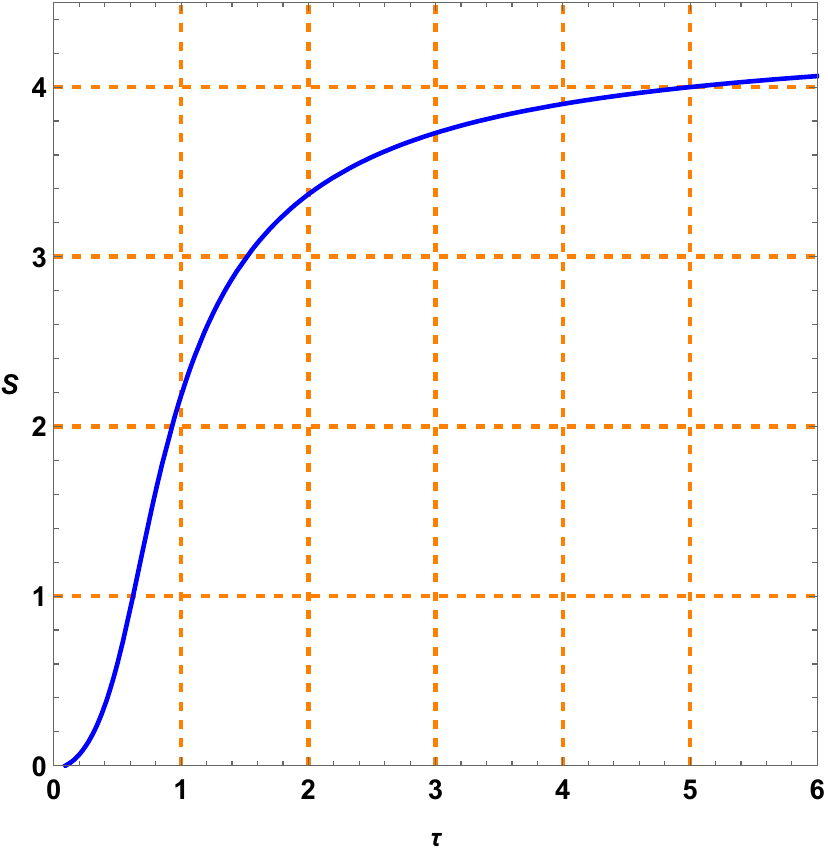}
		\caption{W=-1}
		\label{15c}
	\end{subfigure}
	
	\caption{ $\tau$ vs $S$ plots for Kerr-Sen-Ads black hole in restricted phase space in fixed $(Q,\Omega,C)$ ensemble. Here $\tau$ is plotted against $S$ for three different $\Omega$ values while keeping $l=1,C=0.2,Q=0.05$ fixed. Figure $\left(a\right)$ shows $\tau$ vs $S$ plot at $\Omega=0.05$,figure $\left(b\right)$ shows $\tau$ vs $S$ plot  at $\Omega=5$ and in figure $\left(c\right)$ shows $\tau$ vs $S$ plot at $\Omega=12$.$W$ denotes topological charge.
	}
	\label{15}
\end{figure}

In figure \ref{14}, the effect of variation of $Q$ on topological charge is shown while keeping $l=0.1,C=1,J=0.02$ constant. In Figures \ref{14a}, \ref{14b} and \ref{14c} respectively, we have set $Q=0.05$, $Q=0.5$, and $Q=3.5$. For $Q=0.05$ and $Q=3.5$, two black hole branches are found in figure \ref{14a} and figure.\ref{14c} with topological charge $0$. On the other hand, only one black hole branch is found with topological charge $1$ in figure \ref{14b} when $Q=0.5$ is kept fixed.\\
Lastly, in figures \ref{15a}, \ref{15b}, and \ref{15c} $\tau$ vs $S$ is plotted to check how $\Omega$ affects topological charge while keeping constant $l,Q,C$ at $l=1,Q=0.05,C=1$. $\Omega$ is equal to 0.01 in figure \ref{15a}, 5 in figure \ref{15b}, and 12 in figure \ref{15c}. For $\Omega=5$ and $\Omega=0.05$, the topological charge is $0$ and for $\Omega=12$, the topological charge is found to be $-1$.
In conclusion, our findings show that the Kerr-Sen-Ads black hole in a fixed $(Q,\Omega,C)$ ensemble, the topological charge is $-1,0$ or $1$ depending upon the values of the thermodynamic parameters $Q,\Omega,C$.\\
\subsection{Fixed (Q,J,$\mu$) Ensemble}
\begin{figure}
	\centering
	\includegraphics[width=8cm,height=8cm]{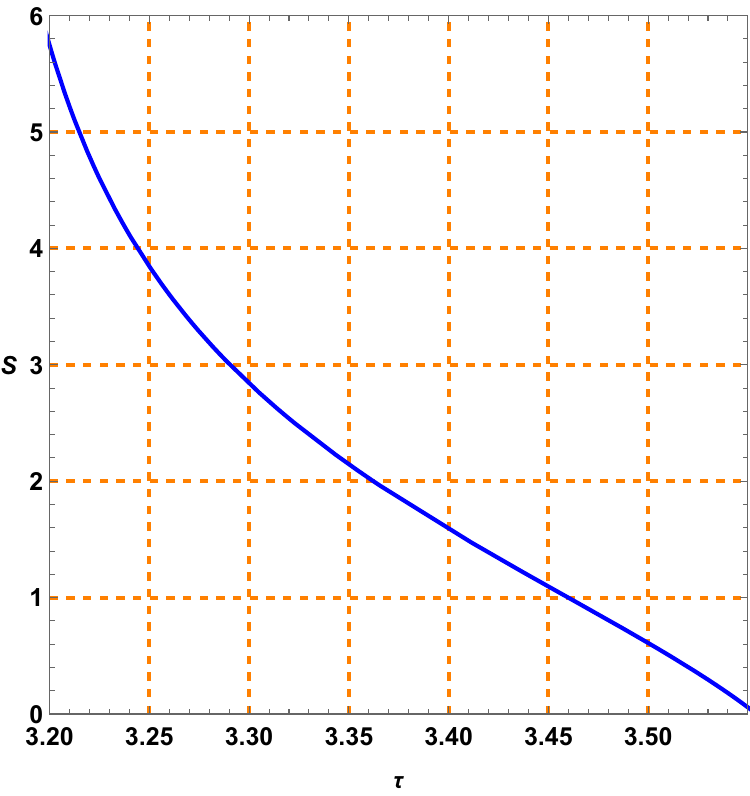}
	\caption{ $\tau$ vs $S$ plots for Kerr-Sen-Ads black hole in restricted phase space in fixed $(Q,J,\mu)$ ensemble. Here $\tau$ is plotted against $S$ while keeping $l=1,\mu=1,Q=1.5$ and $J=1.2$ fixed. The topological charge is found to be $1$.
	}
	\label{16}
\end{figure}

In this ensemble, the chemical potential $\mu$, charge $Q$ and angular momentum $J$ are kept fixed. The chemical potential $\mu$ is given by :
\begin{equation}
	\mu=\frac{S \left(\pi ^2 C^2 \left(4 \pi ^2 J^2+S^2\right)-S^2 \left(2 \pi ^2 Q^2+S^2\right)\right)}{4 \pi ^{3/2} l (C S)^{3/2} \sqrt{\pi  C+S} \sqrt{\pi  C \left(4 \pi ^2 J^2+S^2\right)+2 \pi ^2 Q^2 S+S^3}}
	\label{mu}
\end{equation}
By solving equation \ref{mu}, we get the expression for $C$ and substituting $C$ in the following expression, newly mass($M_{\mu}$) in this ensemble as:
\begin{equation}
	M_{\mu}=M-C \mu
\end{equation}
The off-shell free energy is computed using:
$$\mathcal{F}=M_{\mu}-S/\tau$$
We compute the expressions for $\phi^S$ and $\tau$ using the same method as in the preceding part.\\
It is seen that, for this ensemble, the topological charge is always $1$ irrespective of the values of the thermodynamic parameters $\mu,Q$ and $J.$ Moreover, only one branch is observed in the entropy $S$ vs $\tau$ plot for all allowed values of the thermodynamic parameters. For example, in figure.\ref{16}, the entropy $S$ vs $\tau$ is plotted for $\mu=1,Q=1.5,J=1.2$, and $l=1$. We observe a single black hole branch with a topological charge equal to $1$.\\
\subsection{Fixed ($\phi,\Omega$,C) ensemble}
\begin{figure}[h]	
	\centering
	\begin{subfigure}{0.3\textwidth}
		\includegraphics[width=\linewidth]{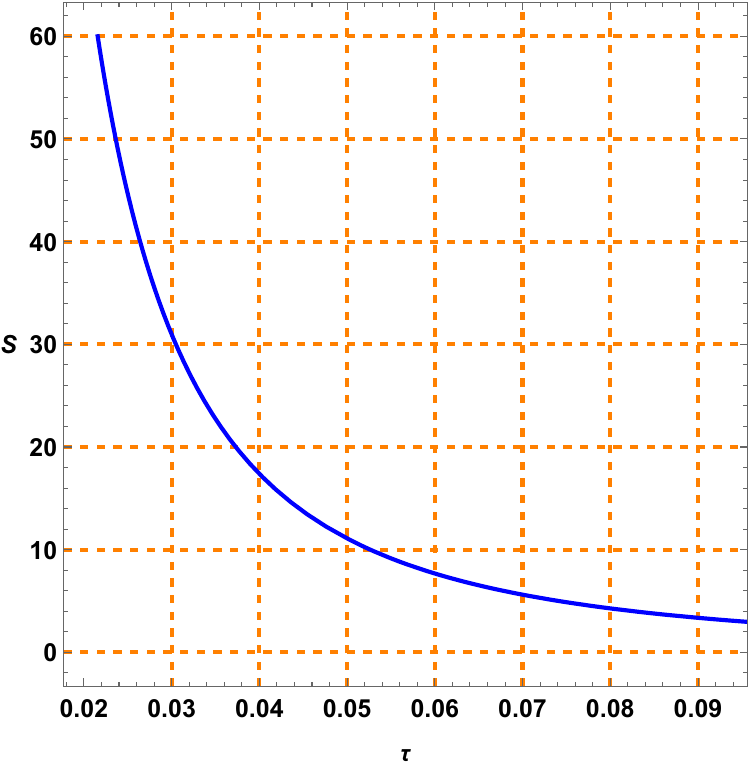}
		\caption{W=1}
		\label{17a}
	\end{subfigure}
	\begin{subfigure}{0.3\textwidth}
		\includegraphics[width=\linewidth]{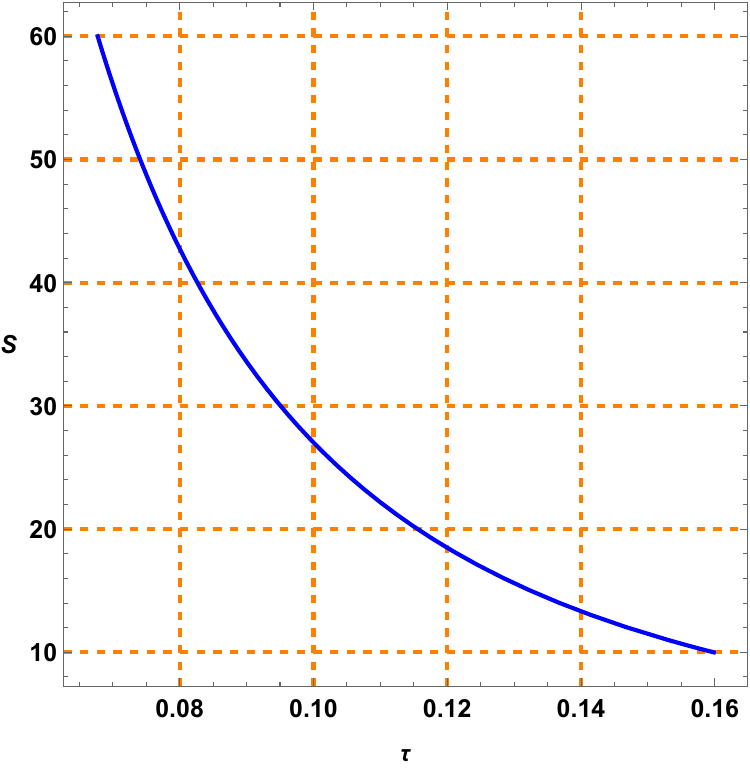}
		\caption{W=1}
		\label{17b}
	\end{subfigure}
	\begin{subfigure}{0.3\textwidth}
	\includegraphics[width=\linewidth]{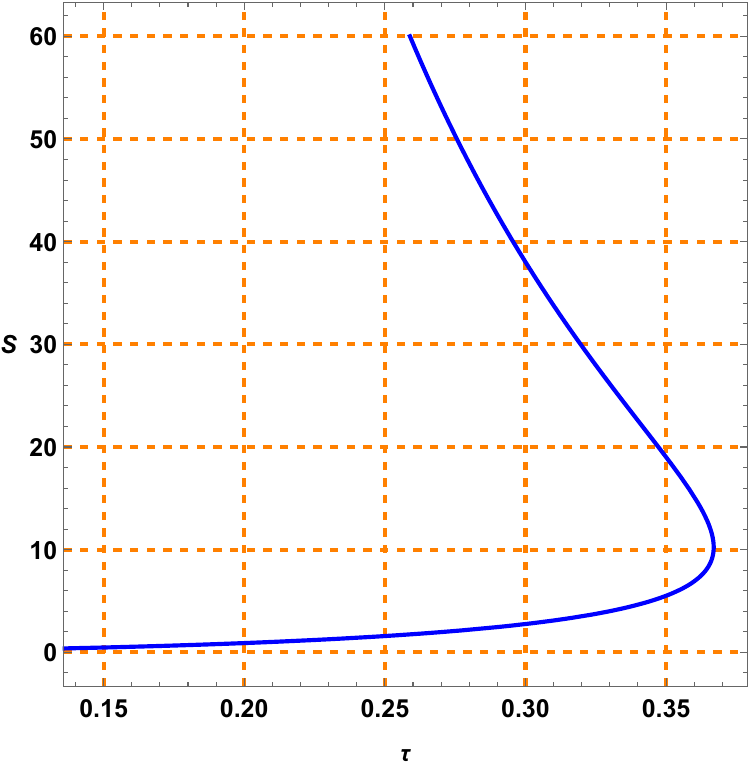}
	\caption{W=0}
	\label{17c}
\end{subfigure}
	
	\caption{  $\tau$ vs $S$ plots for Kerr-Sen-Ads black hole in restricted phase space in fixed $(\phi,\Omega,C)$ ensemble. Here $\tau$ is plotted against $S$ for three different $C$ values while keeping $l=0.1,\phi=3.5,\Omega=1.2$ fixed. Figure $\left(a\right)$ shows $\tau$ vs $S$ plot at $C=0.05$,figure $\left(b\right)$ shows $\tau$ vs $S$ plot  at $C=0.5$ and in figure $\left(c\right)$ shows the same at $C=10$.$W$ denotes topological charge.
	}
	\label{17}
\end{figure}

	The next ensemble we consider is the fixed $(\Omega,\phi,C)$ ensemble. In this ensemble angular frequency $\Omega$ and electric potential $\phi$ are kept fixed. First, We substitute J from equation \ref{omega3} in the expression for mass in equation \ref{mass} as follows:
\begin{equation}
	M_J=\frac{\sqrt{S (\pi  C+S)^2 \left(\pi  C S+2 \pi ^2 Q^2+S^2\right)}}{2 \pi ^{3/2} l \sqrt{C S \left(\pi  C-l^2 S \Omega ^2+S\right)}}
	\label{mj}	
\end{equation}
Now equation \ref{mj} becomes independent of variable $J$. Next we find out $\phi$ from equation \ref{mj} as follows:
\begin{equation}
	\phi=\frac{\partial M_J}{\partial Q}=\frac{\sqrt{\pi } Q \sqrt{S (\pi  C+S)^2 \left(\pi  C S+2 \pi ^2 Q^2+S^2\right)}}{l \sqrt{C S \left(\pi  C S+2 \pi ^2 Q^2+S^2\right) \left(\pi  C-l^2 S \Omega ^2+S\right)}}
	\label{phi}
\end{equation}
Accordingly, $Q$ is given by equation \ref{phi} as:
$$q=\frac{\sqrt{C} l \sqrt{S} \phi  \sqrt{\pi  C+S} \sqrt{\pi  C-l^2 S \Omega ^2+S}}{\sqrt{\pi } \sqrt{-\pi ^2 C^2 \left(2 l^2 \phi ^2-1\right)+2 \pi  C S \left(l^4 \Omega ^2 \phi ^2-l^2 \phi ^2+1\right)+S^2}}$$
Finally, we substitute $Q$ in the following expression for the modified mass in fixed ($\Omega$,$\Phi$) ensemble, which is written as :
\begin{equation}
	\tilde{M}=M-q \phi-J \Omega
	\label{mOP}
\end{equation}
Now $\tilde{M}$ is a function of $\phi,\Omega,C$ and $l.$ Using equation \ref{mOP}, $\mathcal{F}$, $\phi^S$ and $\tau$ are constructed following the standard procedure as shown in the previous sections. \\
We plot $\tau$ vs $S$ curve for different values of $C$ as shown in figure \ref{17} keeping $\Omega=1.2,\phi=3.5,l=0.1$ constant. In figure \ref{17a} and figure \ref{17b}, with $C=0.05$ and $C=0.5$ respectively, one black hole branch with topological charge $W=1$ is observed. At $C=10$, in figure \ref{17c} two black hole branches and topological charge $0$ are found.\\

\begin{figure}[h]	
	\centering
	\begin{subfigure}{0.3\textwidth}
		\includegraphics[width=\linewidth]{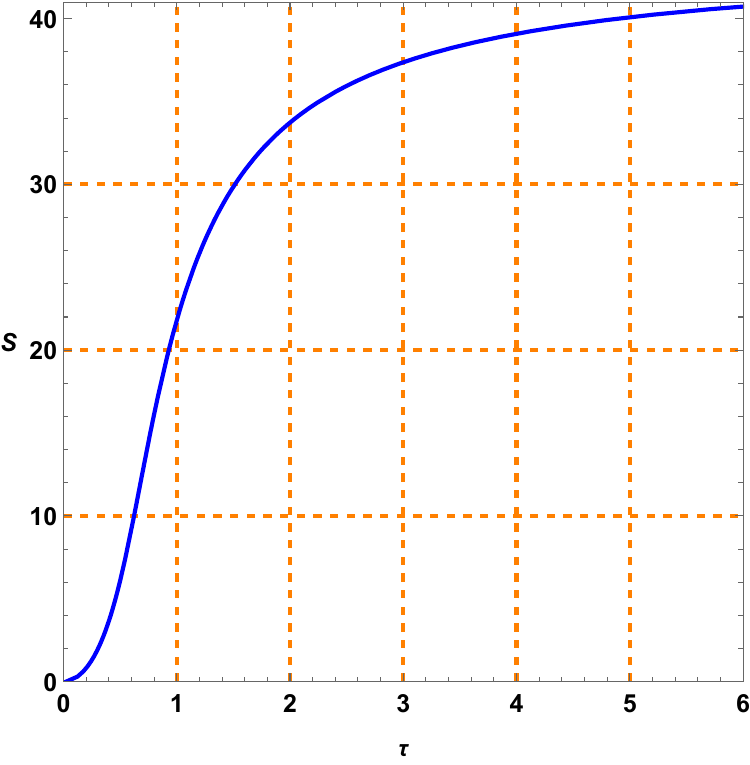}
		\caption{W=-1}
		\label{18a}
	\end{subfigure}
	\begin{subfigure}{0.3\textwidth}
		\includegraphics[width=\linewidth]{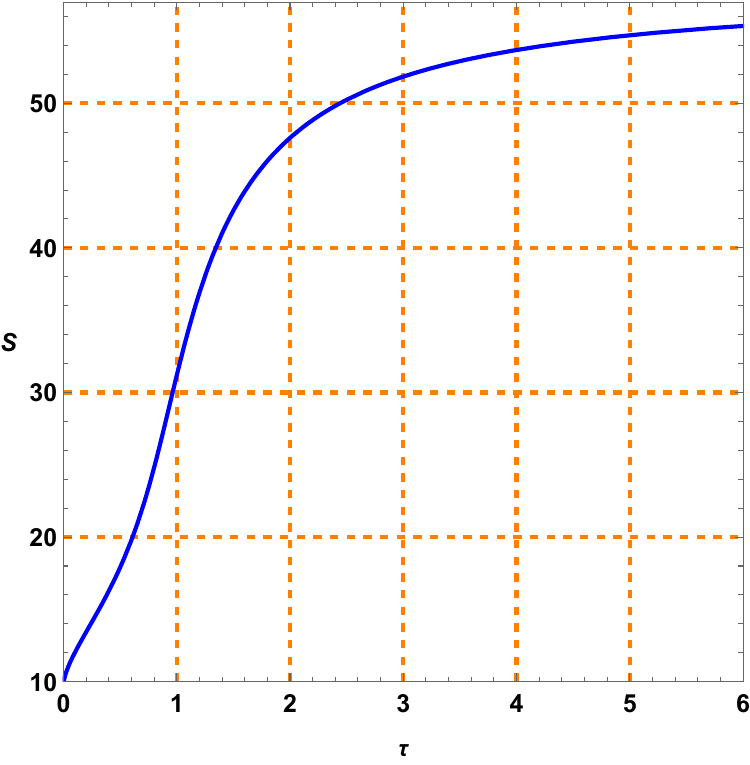}
		\caption{W=-1}
		\label{18b}
	\end{subfigure}
	\begin{subfigure}{0.3\textwidth}
	\includegraphics[width=\linewidth]{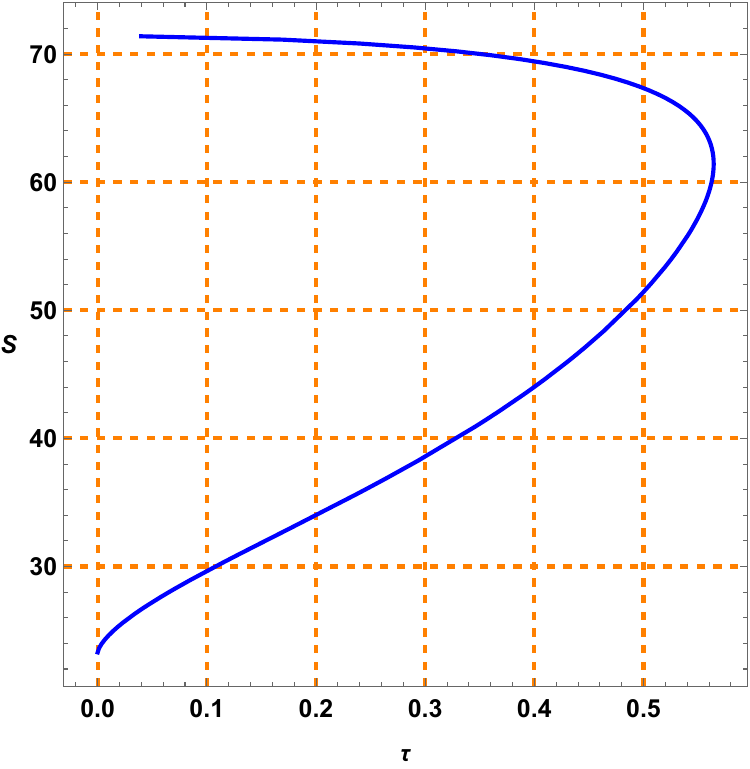}
	\caption{W=0}
	\label{18c}
\end{subfigure}
	
	\caption{  $\tau$ vs $S$ plots for Kerr-Sen-Ads black hole in restricted phase space in fixed $(\phi,\Omega,C)$ ensemble. Here $\tau$ is plotted against $S$ for three different $\phi$ values while keeping $l=0.1,C=10, \Omega=12$ fixed. Figure $\left(a\right)$ shows $\tau$ vs $S$ plot at $\phi=1$,figure $\left(b\right)$ shows $\tau$ vs $S$ plot  at $\phi=10$ and in figure $\left(c\right)$ shows the same at $\phi=15$.$W$ denotes topological charge.
	}
	\label{18}
\end{figure}

Next we investigate the topological charge with reference to a variation in the potential $\phi$ while keeping $C=10,\Omega=12,l=0.1$ fixed. In figure \ref{18a} and \ref{18b} we set $\phi=1$ and $\phi=10$ respectively where we find one black hole branch and topological charge $W=-1$. In the latter case, where $\phi=15$, we find two black hole branches with topological charge $0.$ \\

\begin{figure}[h]	
	\centering
	\begin{subfigure}{0.3\textwidth}
		\includegraphics[width=\linewidth]{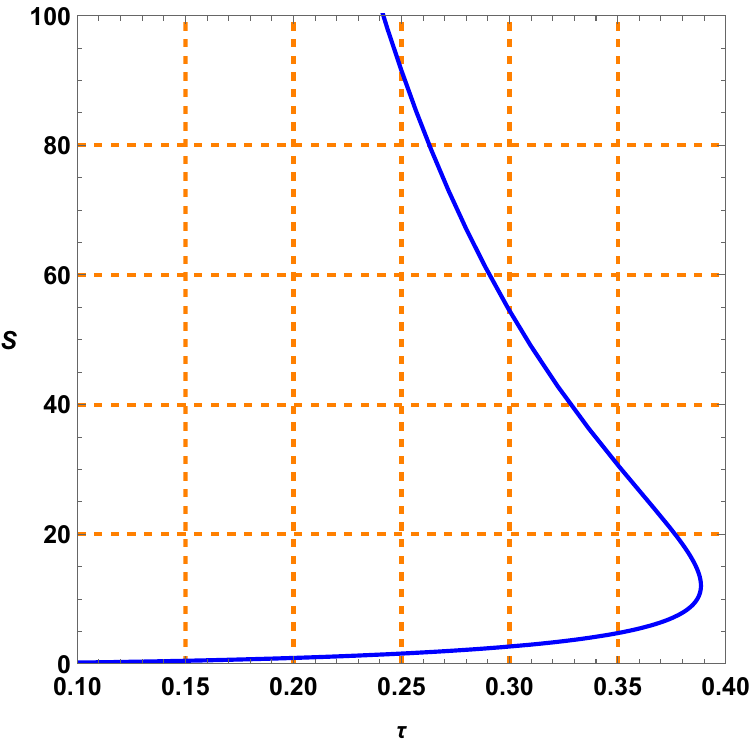}
		\caption{W=0}
		\label{19a}
	\end{subfigure}
	\begin{subfigure}{0.3\textwidth}
		\includegraphics[width=\linewidth]{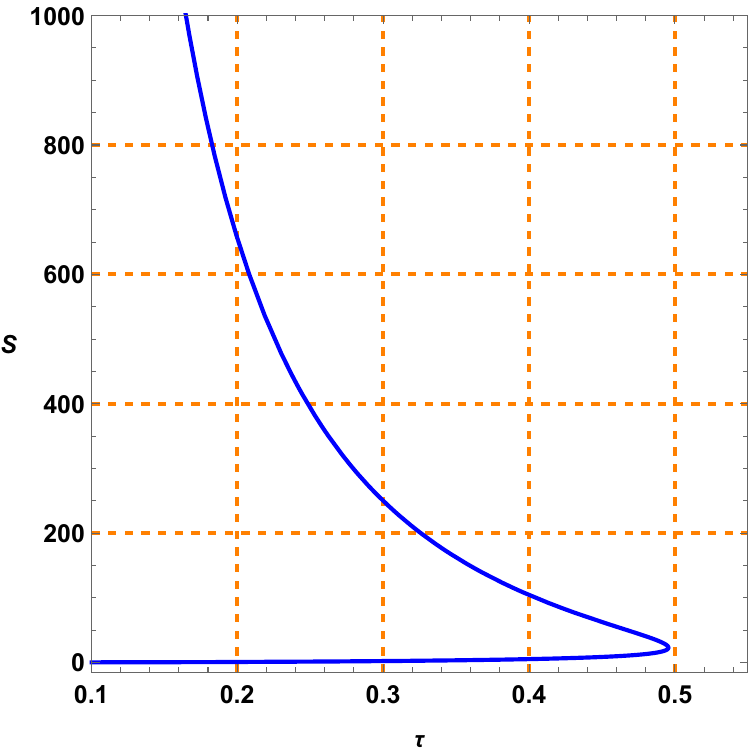}
		\caption{W=0}
		\label{19b}
	\end{subfigure}
	\begin{subfigure}{0.3\textwidth}
	\includegraphics[width=\linewidth]{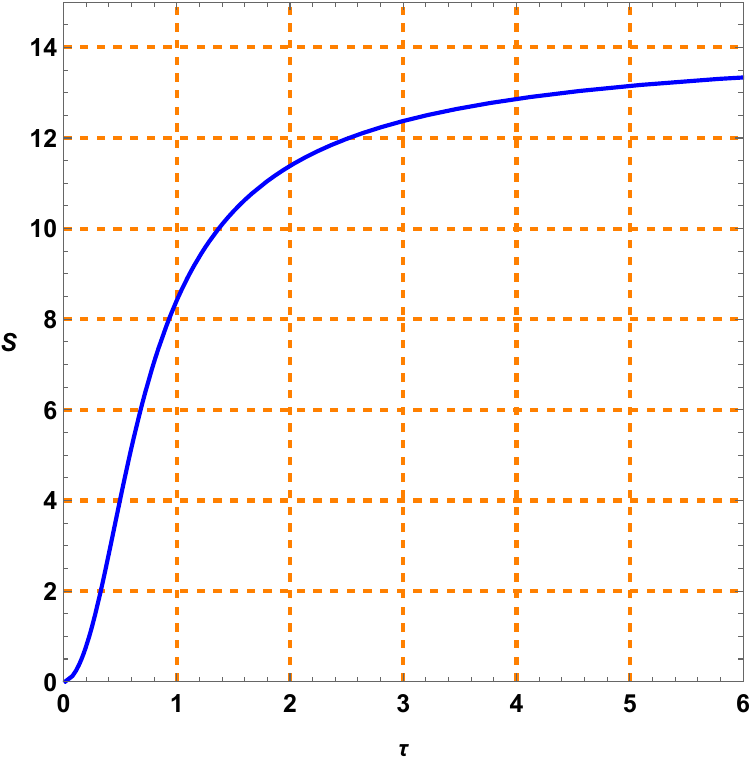}
	\caption{W=-1}
	\label{19c}
\end{subfigure}
	\caption{  $\tau$ vs $S$ plots for Kerr-Sen-Ads black hole in restricted phase space in fixed $(\phi,\Omega,C)$ ensemble. Here $\tau$ is plotted against $S$ for three different $\Omega$ values while keeping $l=0.1,C=1,\phi=0.01$ fixed. Figure $\left(a\right)$ shows $\tau$ vs $S$ plot at $\Omega=5$,figure $\left(b\right)$ shows $\tau$ vs $S$ plot  at $\Omega=10$ and in figure $\left(c\right)$ shows the same at $\Omega=15$.$W$ denotes topological charge.
	}
	\label{19}
\end{figure}

Finally,in figures \ref{19}, we fix $\phi=0.01$,$C=10,l=0.1$ and vary $\Omega.$ In figure \ref{19a} and figure \ref{19b}, setting $\Omega=5$ and $\Omega=10$ respectively, we get two black hole branches and topological charge $W=0.$ In figure \ref{19c}, with $\Omega=15$ a single black hole branch with topological charge $W=-1$ is observed.\\ 

So to summarize, the topological charge in fixed $(\phi,\Omega,C)$ ensemble is $-1$,$0$ or $+1$ depending on all the thermodynamic parameters $C,\Omega$ and $\phi.$\\

\begin{figure}[h!t]
	\centering
	\includegraphics[width=10cm,height=9cm]{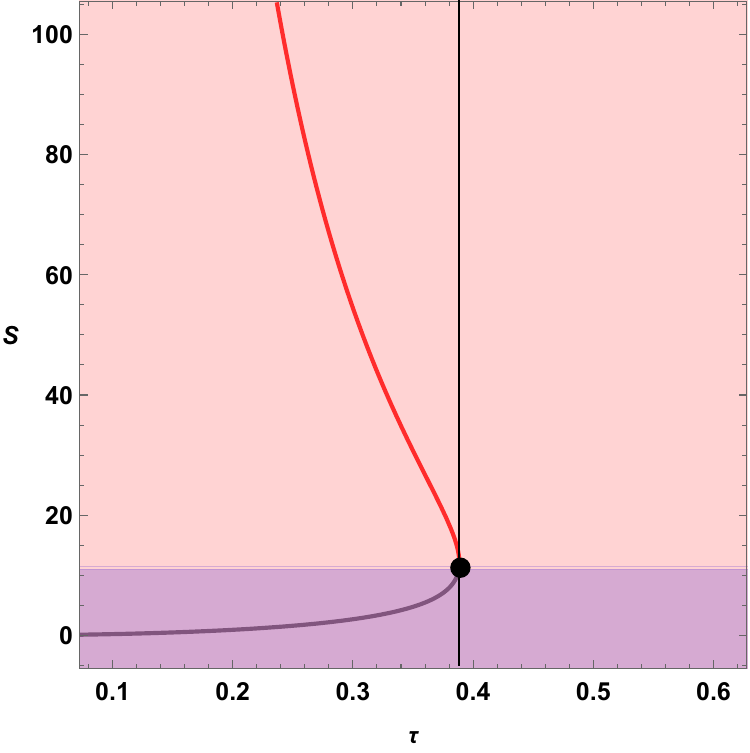}
	\caption{$\tau$ vs $S$ plot for Kerr Sen Ads  black hole at  $C=10,\phi=0.01,\Omega=5$ and $l=0.1$. The red and the black-coloured solid represent the large and small black hole branches respectively.}
	\label{20}
\end{figure}

\begin{figure}[h!t]
	\centering
	\begin{subfigure}{0.4\textwidth}
		\includegraphics[width=\linewidth]{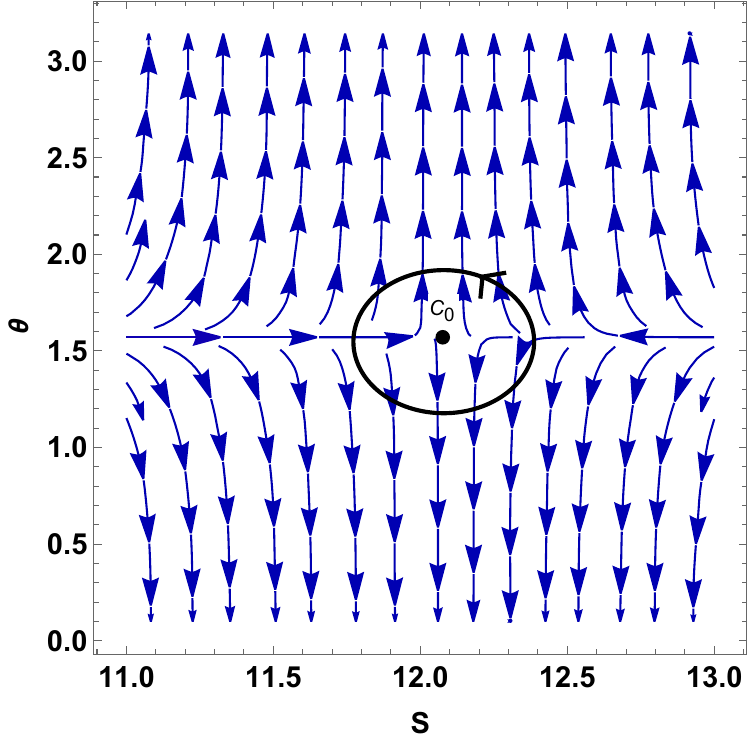}
		\caption{}
		\label{21a}
	\end{subfigure}
	\begin{subfigure}{0.4\textwidth}
		\includegraphics[width=\linewidth]{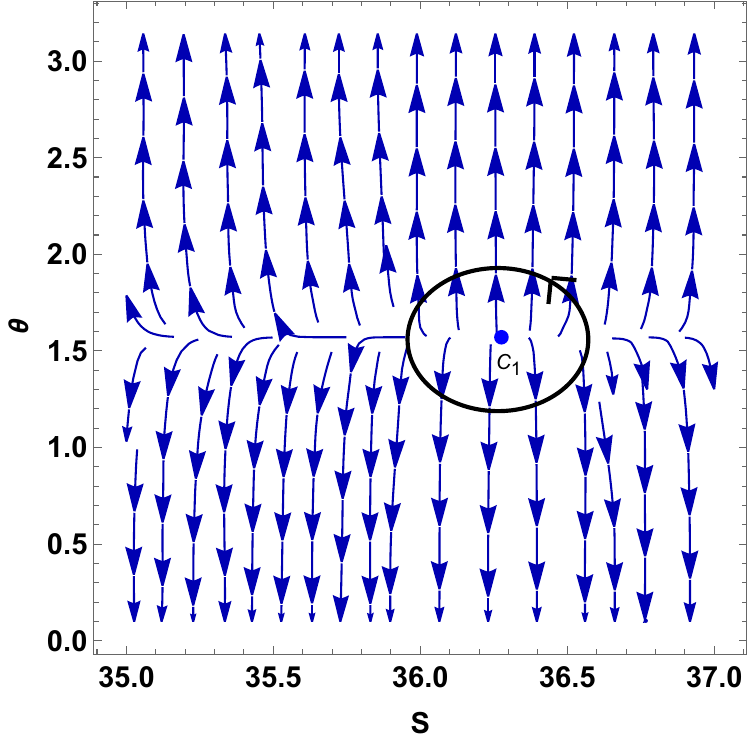}
		\caption{}
		\label{21b}
	\end{subfigure}
	\caption{A portion of the vector field $\phi$ on $S-\theta$ plane. The black dot of Figure $(a)$ represents Davies point and the blue dot in Figure $(b)$ represents Hawking Page phase transition point.}
	\label{21}
\end{figure}

\begin{figure}[h]	
	\centering
	\begin{subfigure}{0.40\textwidth}
		\includegraphics[width=\linewidth]{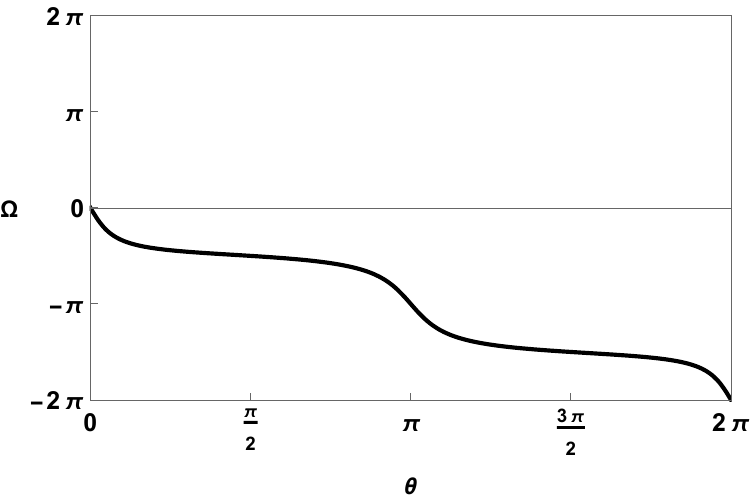}
		\caption{}
		\label{}
	\end{subfigure}
	\hspace{0.5cm}
	\begin{subfigure}{0.40\textwidth}
		\includegraphics[width=\linewidth]{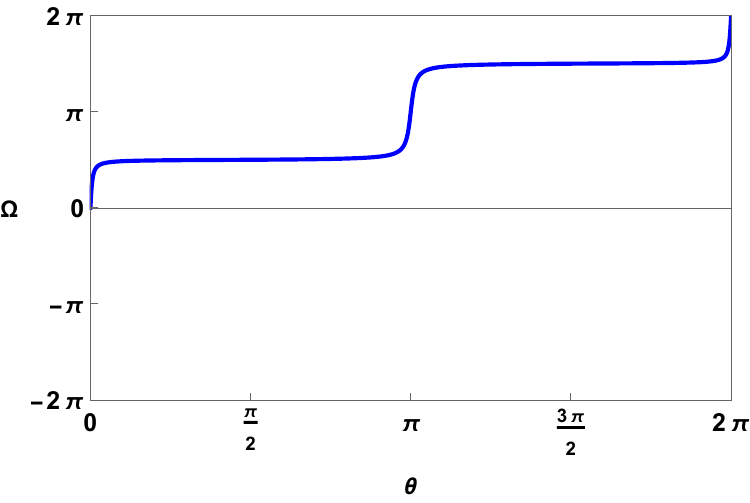}
		\caption{}
		\label{}
	\end{subfigure}
	
	\caption{Kerr Sen AdS black hole in fixed ($\phi$,$\Omega$,C) ensemble: Figure (a) represents topological number calculation  for Davies point around the $C_0$ contour in Figure.\ref{2} and Figure (b) represents the topological number calculation  for Hawking-Page phase transition point around the $C_1$ contour
	}
	\label{21.1}
\end{figure}

Like fixed ($Q,\Omega,C$) ensemble, here also we find a class of black holes, where  topological charge is found to be zero. Here also both Hawking-Page transition and Davies type of phase transition are observed. For example, taking $C=10,\phi=0.01,\Omega=5$ and $l=0.1$ we observe two black hole branch as shown in figure.\ref{20}
Using the following equation we construct the vector field 
\begin{equation}
	\phi=\left(\frac{1}{S}\frac{\partial F^2}{\partial S},-cot\theta csc\theta\right)
	\label{new}
\end{equation}
The zero point of the vector field is found to be $(12.0763,\frac{\pi}{2})$ and  $(36.276,\frac{\pi}{2})$. The vector plot is shown in the figure.\ref{21}
The black dot ($S= 12.0763$) in figure \ref{21a} is a Davies transition point and the blue dot ($S=36.276$) is a Hawking-Page phase transition point. We have confirmed that, these zero points exactly matches the critical points of the Hawking-page and Davies type phase transition. The topological charge is calculated by constructing a contour $C$ around each zero point($C_0$ and $C_1$ as shown in figure \ref{21}). As expected, the topological charge of the Hawking page phase transition is $+1$ and that of Davis point is found to be $-1$ as shown in figure.\ref{21.1}.\\

We found the topological class $W=0$ in the Kerr-Sen black hole (non-AdS), similar to what we observed in the $(Q,\Omega,C)$ and $(\phi,\Omega,C)$ ensembles within the RPST framework. Despite this shared topological class of 0, the local topology differs significantly between the two cases. In the Kerr-Sen-AdS black hole, using the RPST formalism, the small black hole is unstable with a winding number of -1, while the large black hole is stable with a winding number of 1. Conversely, for the Kerr-Sen black hole without the AdS boundary in regular thermodynamic space, the situation is reversed: the small black hole is stable with a winding number of 1, and the large black hole is unstable with a winding number of -1.
Moreover, in the Kerr-Sen-AdS black hole, we observe both Hawking-Page and Davies-type phase transitions within the RPST formalism. In contrast, for the Kerr-Sen black hole without the AdS boundary, we find only the Davies-type phase transition.
\section{Conclusion}
	In this study, the thermodynamic topology of the Kerr-Sen-Ads black hole in restricted phase space is analyzed. For our analysis, we have considered five ensembles of Kerr-Sen-Ads black holes in restricted phase space: fixed $(Q, J, C)$, fixed $(\phi, J, C)$, fixed $(Q,\Omega, C)$, fixed $(Q, J, \mu)$, and fixed $(\phi,\Omega, C)$. We studied the local and global topology in each ensemble by computing the topological number. In the fixed $(Q, J, C)$, fixed $(\phi, J, C)$, and fixed $(Q, J, \mu)$ ensembles, we found a topological charge of $+1$. In the fixed $(Q,\Omega, C)$ and fixed $(\phi, \Omega, C)$ ensembles, we found topological charges of $-1$, $0$, and $+1$.  The results are summarized below. 
	\begin{center}
		\begin{tabular}{c c c c c} 
			\hline
			\hline
			& Ensembles \hspace{0.4cm}& Topological Charge\hspace{0.3cm} & Annihilation Point\hspace{0.3cm} & Generation Point \\
			\hline
			\hline
			\vspace{-0.2cm}\\
			&Fixed ($Q, J, C$)  &1&1 or 0 & 1 or 0\\
			&Fixed ($\phi, J, C$) &1  & 1 or 0 & 1 or 0\\
			&Fixed ($Q, \Omega, C$) &-1, 0 or 1 & 0 &1 or 0 \\
			&Fixed ($Q, J, \mu$) & 1 & 0 &0 \\
			&Fixed ($\phi, \Omega, C$) &-1, 0 or 1 & 0 & 1 or 0 \\
			\hline
			\hline
			\label{table1}
		\end{tabular}
	\end{center}
Notably, within ensembles where the topological charge is determined to be $0$, the occurrence of both Hawking-Page and Davies-type phase transitions is observed. We have shown that these distinct types of phase transitions can be comprehensively investigated by employing a common vector field. Furthermore, both types of phase transitions can be distinguished by the associated topological charges. The topological charge linked with Davies-type and Hawking-Page phase transitions is found to be $-1$ and $+1$, respectively.\\

\textbf{Note :} It is important to mention that three more possible ensembles exist in this system: fixed ($\phi, J, \mu$), fixed ($Q, \Omega, \mu$), and fixed ($\phi, \Omega, \mu$). However, due to mathematical complexities, a closed-form solution for the black hole mass could not be obtained in these ensembles, and thus topological analysis was not performed for these cases.
	\section{Acknowledgments}
	BH would like to thank DST-INSPIRE, Ministry of Science and Technology fellowship program, Govt. of India for awarding the DST/INSPIRE Fellowship[IF220255] for financial support.

\end{document}